\documentclass[letterpaper,sn-nature]{sn-jnl}
\usepackage[version=3]{mhchem} 

\usepackage[normalem]{ulem}
\usepackage{cancel}
\usepackage{amsthm}
\newtheorem{theorem}{Theorem}
\usepackage[export]{adjustbox}
\usepackage{xcolor}
\definecolor{C0}{HTML}{1F77B4}
\definecolor{C1}{HTML}{FF7F0E}
\definecolor{C2}{HTML}{2ca02c}
\definecolor{C3}{HTML}{d62728}
\definecolor{C4}{HTML}{9467bd}
\definecolor{C5}{HTML}{8c564b}

\def\usetodonotes{} 
\ifdefined\usetodonotes
\usepackage{todonotes}
\usepackage{changes}

\usepackage{setspace}

\fi

\usepackage{tabularx}
\usepackage{upgreek}
\usepackage{tikz}
\usepackage{soul}
\usepackage{tcolorbox}
\usepackage[scr=esstix,cal=boondox]{mathalfa}
\usepackage{booktabs}
\usepackage{graphicx}
\usepackage{dcolumn}
\usepackage{bm}

\usepackage{siunitx}
\usepackage[skip=0pt]{caption}
\usepackage{subfig}
\usepackage{float}
\usepackage{amsmath}
\usepackage{amsmath}
\usepackage{amsfonts}
\DeclareMathOperator{\curl}{curl}
\newcommand{\Fe}{\bm F}
\newcommand{\jump}[1]{[\![ #1 ]\!]}

\newcommand{\InvFe}{\bm F^{-1}}
\newcommand{\angstrom}{\textup{\AA}}
\newcommand{\bol}[1]{\overline{\bm{#1}}}

\newcommand{\bul}[1]{\underline{\bm #1}}

\usepackage{natbib}
\setcitestyle{square, numbers, comma, sort&compress, super}
\usepackage{algpseudocode}
\usepackage{algorithm}
\usepackage{latexsym}

\hypersetup{ 
    colorlinks=true,       
    linkcolor=red,          
   citecolor=blue,        
    filecolor=magenta,      
    urlcolor=cyan           
}
\usepackage{cleveref}

\newcommand{\psubref}[1]{\protect\subref{#1}}

\begin{document}

\title[]{Inverse design of heterodeformations for strain soliton networks in bilayer 2D materials}


\author[1,2]{\fnm{Md Tusher} \sur{Ahmed}}\email{mdtusher.ahmed@utrgv.edu}

\author*[2]{\fnm{Nikhil Chandra} \sur{Admal}}\email{admal@illinois.edu}

\affil[1]{\orgdiv{Institute for Advanced Manufacturing}, \orgname{The University of Texas Rio Grande Valley}, \orgaddress{\street{501 N Sugar Rd}, \city{Edinburg}, \postcode{78539}, \state{Texas}, \country{USA}}}

\affil[2]{\orgdiv{Department of Mechanical Science and Engineering}, \orgname{University of Illinois, Urbana-Champaign}, \orgaddress{\street{1206 W Green St}, \city{Urbana}, \postcode{61801}, \state{Illinois}, \country{USA}}}

\abstract{
Strain soliton networks strongly influence the structural and electronic properties of heterodeformed bilayer systems, yet their design remains challenging due to the high dimensionality of heterodeformation space and the absence of a direct map between deformation and network geometry. In this work, we introduce a geometric framework that establishes a one-to-one mapping between heterodeformations and the geometry of the strain soliton network expressed as line vector–Burgers vector pairs. The admissible networks are constrained by topology dictated by the generalized stacking fault energy landscape. We show that the moir\'e Bravais lattice, corresponding to a uniform heterodeformation, alone is insufficient to characterize the interface: distinct heterodeformations can share identical moir\'e Bravais lattices while producing different soliton networks, reflecting an inherent many-to-one mapping when only translational symmetry is considered. In contrast, the soliton network encodes the full multilattice geometry of the interface, including topology and connectivity, which are not captured by the moir\'e Bravais lattice alone. The proposed framework enables the direct construction of heterodeformations from target networks, providing a systematic route for inverse design of moir\'e interfaces beyond conventional twist-based approaches.

}
\maketitle
\section{Introduction}
\label{sec:intro}
In the era of micro- and nano-electromechanical devices, 2D materials have garnered significant attention from the scientific community due to their superior thermal, mechanical, and electronic properties \cite{geim2007rise,castro2009electronic,ferrari2015science}. Particularly, the observation of flat bands at magic angle twisted bilayer two-dimensional (2D) materials gives rise to interesting quantum properties such as quantum hall effect, Mott insulating effect, superconductivity, etc. \cite{po2018origin,novoselov2006unconventional}, which makes the use of heterodeformed\footnote{Heterodeformation is an umbrella term used to refer to relative twists and relative strains.} bilayer 2D materials versatile for numerous applications \cite{buch2018superlubricity,li2017superlubricity}. A rigid twisted and/or heterostrained bilayer 2D material undergoes structural reconstruction where atoms are displaced from their positions in the rigidly heterodeformed configuration \cite{popov_chain_2011,Koshino_2017}. This reconstruction is mediated by the formation of an array of interface dislocations, also referred to as \emph{strain solitons} \cite{Koshino_2017,harley_disloc,Annevelink_2020,ahmed2024bicrystallography,ahmed2025quantifying}. These solitons are characterized by their Burgers vector and line direction \cite{srolovitz2015}. In twisted bilayer graphene, they typically organize into straight triangular networks \cite{kazmierczak2021strain}, whereas heterostrained bilayers may exhibit distorted triangular \cite{kazmierczak2021strain,van2023rotational} or even spiral networks \cite{mesple2023giant}. The structure of each soliton depends on the alignment between its Burgers vector and line direction, imparting screw-like or edge-like character \cite{cazeaux2023relaxation,zhang2024impact}.

\begin{figure}[H]
    \centering
    \subfloat[]
    {
\includegraphics[width=0.3\textwidth]{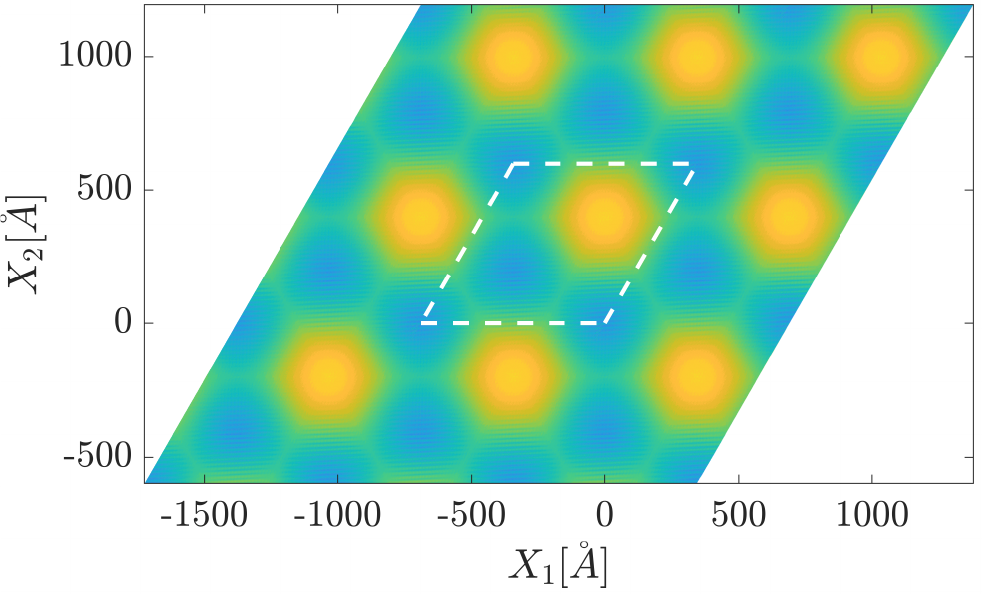}
        \label{fig:triangle_mixed_unrel}
    }
    \subfloat[]
    {
\includegraphics[width=0.28\linewidth]{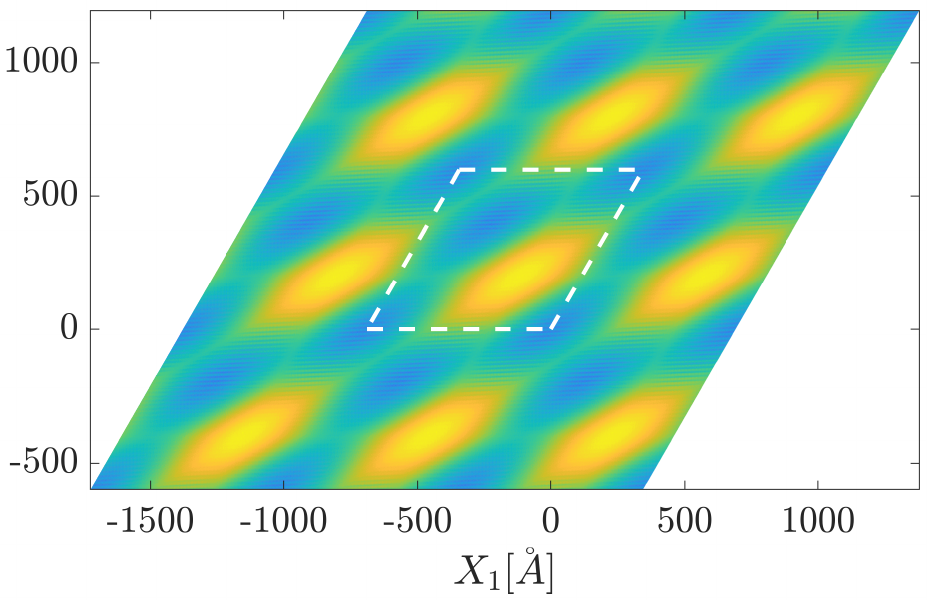}
    \label{fig:triangle_uneq_unrel}
    }
    \subfloat[]
    {
\includegraphics[width=0.31\linewidth]{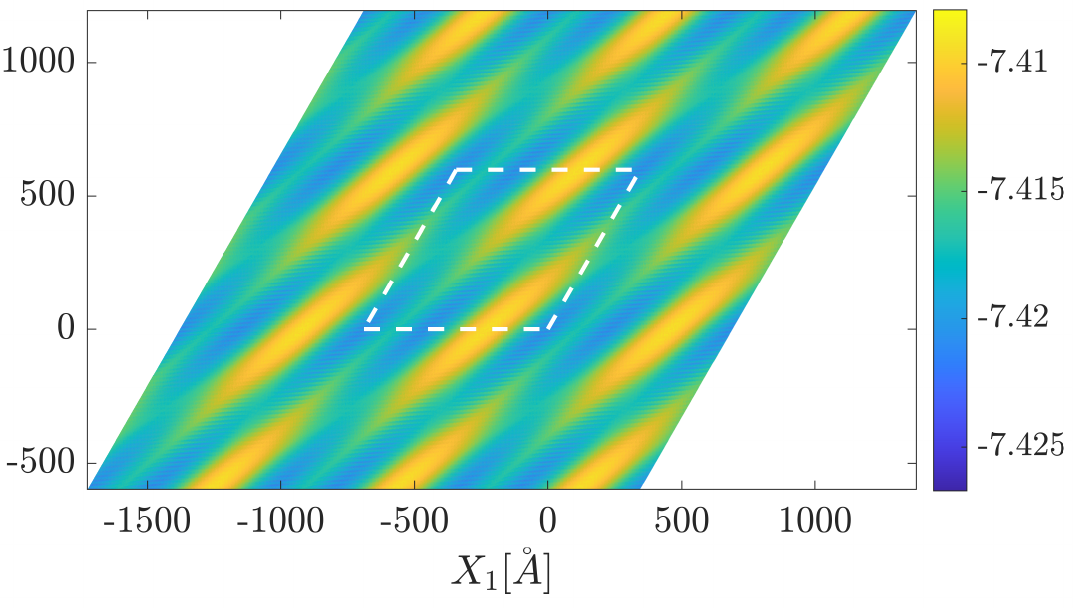}
    \label{fig:mixed_hetero_unrel}
    }
    \caption{Three heterodeformations of bilayer graphene, resulting in identical 2D moir\'e Bravais lattices. The three configurations, however, have distinct point group symmetries. The colors represent energy per atom in [$\si{\electronvolt}$].}
\label{fig:disloc_degenracy_unrel}
\end{figure}

The geometry of the soliton network is not merely a structural detail. In continuum models of flat band physics \cite{Macdonald_2011,tang2020simulation,carr2019exact,zou2018band, koshino2020effective}, the effective interlayer coupling is assumed to be periodic with respect to the \emph{2D moir\'e Bravais lattice} generated by the imposed heterodeformation. In other words, the 2D moir\'e Bravais lattice determines the translational symmetry of the interlayer coupling. It does not specify the point group symmetry of the interlayer coupling. Only when the moir\'e is viewed in its entirety, i.e., the moir\'e Bravais lattice along with the atoms of both layers interpreted as its basis atoms, do the internal shifts determine the point group symmetry of the interlayer coupling. We will use the term \emph{moir\'e lattice} to refer to the entire multilattice. As an example, \Cref{fig:disloc_degenracy_unrel} shows three distinct heterodeformed configurations of bilayer graphene that share a common 2D moir\'e Bravais lattice, whose primitive unit cell is shown by a white dashed parallelogram. Although the three moir\'e lattices have identical translational symmetries, they have distinct point group symmetries. \emph{
    Consequently, two heterodeformations producing identical 2D moir\'e Bravais lattices may nevertheless yield continuum Hamiltonians with distinct point group symmetries and band structures
}. Moreover, anisotropic band structure observed in distorted, heterostrained bilayer configurations can be attributed to the strain-induced distortion of the moir\'e lattice and the resulting modification of the moir\'e Brillouin zone \cite{sequeira2024manipulating}.

Structural relaxation of the three configurations in \Cref{fig:disloc_degenracy_unrel} yields distinct soliton networks that inherit the translational and point group symmetries of the moir\'e lattice. In addition to the band structure, the soliton network also plays an important role in superlubricity and ferroelectricity. The frictional response of a heterodeformed bilayer is determined by the mobility of interface dislocations \cite{ahmed2025quantifying}, which depends on the Burgers vector, line direction, and network connectivity. In ferroelectric systems like bilayer hBN, domain transitions under applied electric field are mediated by the deformation (bending/swirling) of interface dislocations \cite{weston2022interfacial,ahmed2025multiscale}, which depends on the geometry of interface dislocations. Therefore, controlling the geometry of the network of strain solitons is critical for the design of electronic and mechanical properties of heterodeformed bilayer materials.

It is well established that heterodeformation controls the characteristic size and anisotropy of moir\'e lattices in bilayer 2D materials. Experiments and simulations show that increasing twist angle or heterostrain modifies domain-wall width and reduces the moir\'e length scale \cite{ishikawa2016interfacial,kazmierczak2021strain,mesple2023giant}, while numerous theoretical studies demonstrate that the moir\'e lattice unit cell can be tuned continuously by adjusting the relative deformation \cite{uchida2014atomic,van2015relaxation,koda2016coincidence,kogl2023moire,cazeaux2023relaxation,escudero2024designing}. In particular, recent works \cite{kogl2023moire,escudero2024designing} treat heterodeformation as a control parameter for generating moir\'e lattices. However, these approaches constitute a \emph{forward-design} rather than an inverse-design paradigm as their input is the imposed heterodeformation (e.g., heterostrain and/or twist), and the output that is usually emphasized is the 2D moir\'e Bravais lattice. 
Since the forward design is a many-to-one mapping---as demonstrated in \Cref{fig:disloc_degenracy_unrel}---it does not fully establish the relationship between the moir\'e and heterodeformation, which is a key requirement for experiments \cite{he2021moire,kapfer2023programming}. Therefore, our goal is to develop an inverse design framework wherein \emph{the input is the target moir\'e---fully described by not only the Bravais lattice vectors but also the soliton network---and the output is the heterodeformation}. We will show that the inverse design is a one-to-one mapping and, therefore, mathematically well-posed. A key component of our inverse design is Smith normal form (SNF) bicrystallography \cite{ADMAL2022}, an algebraic tool for analyzing the bicrystallography of crystal interfaces.

We first introduce the geometric framework and notation, followed by the inverse design formulation and its application to representative bilayer systems. All atomistic configurations are plotted using the Open Visualization Tool (OVITO) \cite{ovito}.

\section{Notation and lattice preliminaries}
\label{sec:notation}
Lowercase bold letters are used to denote vectors while uppercase bold letters are used to denote second-order
tensors and matrices, unless stated otherwise. The symbol $\cdot$ denotes the inner product of two vectors.
The \emph{Cartesian coordinates} of a vector $\bm x$ are represented as a column matrix $\bm x = (x_1,\dots,x_n)^{\rm T}$, where $\rm T$ denotes transpose. The curl of a second-order tensor $\bm T$ is a tensor whose components are given by $[\mathrm{curl}(\bm T)]_{ij}=\epsilon_{ipq} T_{jq,p}$. The outer product of two vectors, denoted by $\bm a \otimes \bm b$, is a tensor with components $[\bm a \otimes \bm b]_{ij}=a_i b_j$. 

An $n$-dimensional \emph{Bravais lattice} is defined as
\begin{equation}
    \mathcal{A}=\left\{\bm x\in \mathbb{R}^n \text{, such that } \bm x=\bm A\bul x\text{ and }\bul x\in \mathbb{Z}^n\right\},
    \label{eqn:lattice}
\end{equation}
where  $\bm A$ is an $n \times n$ \textit{structure matrix} of the lattice, whose columns are its primitive lattice vectors. In other words, $\mathcal A$ is the integer linear combinations of its primitive lattice vectors. The volume of a primitive unit cell of $\mathcal A$ is equal to $\det \bm A$. We use an uppercase calligraphic letter to denote a lattice; the same letter in bold denotes its structure matrix. The \emph{lattice coordinates} of a $\bm x \in \mathbb R^n$ are denoted by $\bul x=(\underline x_1,\dots,\underline x_n)^{\rm T}$, and given by $\bul x = \bm A^{-1} \bm x$. It follows from \cref{eqn:lattice} that if $\bm x$ is a lattice vector, its lattice coordinates are integers. If lattice $\mathcal A$ is subjected to a uniform deformation $\bm x'=\bm F \bm x$ using a constant deformation gradient $\bm F$, the resulting deformed lattice is denoted by $\bm F \mathcal A$.

Reciprocal lattice vectors of lattice $\mathcal A$ are the lattice vectors of its dual lattice given by
\begin{equation}
    \mathcal{A}^*=\left\{\bm y \in \mathbb{R}^n \text{, such that } \bm y=\bm A^{-\rm T} \bol  y\text{ and }\bol y\in \mathbb{Z}^n\right\},
    \label{eqn:dual_lattice}
\end{equation}
whose structure matrix is $\bm A^{-\rm T}$. The dual lattice coordinates of $\bm y$ are denoted by $\bol y$. From \cref{eqn:lattice,eqn:dual_lattice}, it follows that if $\bm x \in \mathcal A$ and $\bm y \in \mathcal A^*$, then $\bm x \cdot \bm y = \bul x \cdot \bol y$ is an integer. We reserve the term $\emph{lattice}$ to represent a generic multilattice formed by a Bravais lattice $\mathcal A$ and a collection of basis atom positions (shift vectors) $S = \{\bm s^1,\dots,\bm s^m\}$, i.e. 
\begin{equation}
    \mathcal A + \{\bm s^\alpha\} : = \{\bm x + \bm s^\alpha \text{, such that }  \bm x \in \mathcal A, \bm s^\alpha \in S\}.
\end{equation}
If $S=\{\bm 0\}$, we recover the Bravais lattice.

\section{Moir\'e superlattice and the geometry of strain solitons}
\label{sec:geom_str}
\begin{figure}[t]
    \subfloat[bilayer graphene]
    {
        \includegraphics[height=0.21\textwidth]{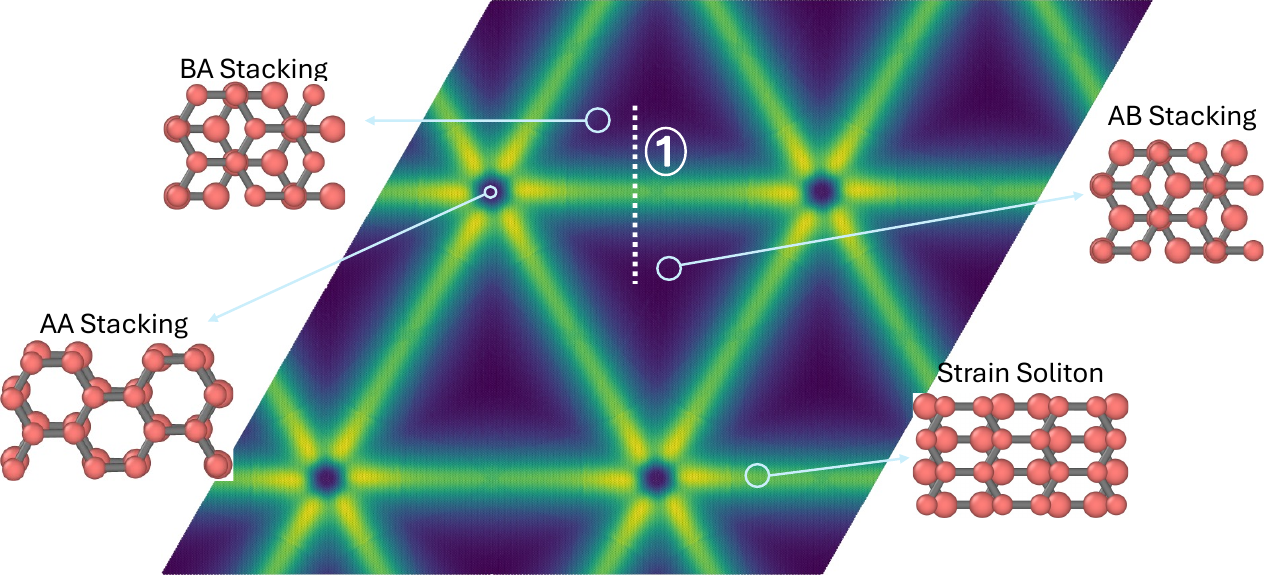}
        \label{fig:tri_net}
    }
    \subfloat[bilayer MoS$_2$]
    {
        \includegraphics[height=0.21\textwidth]{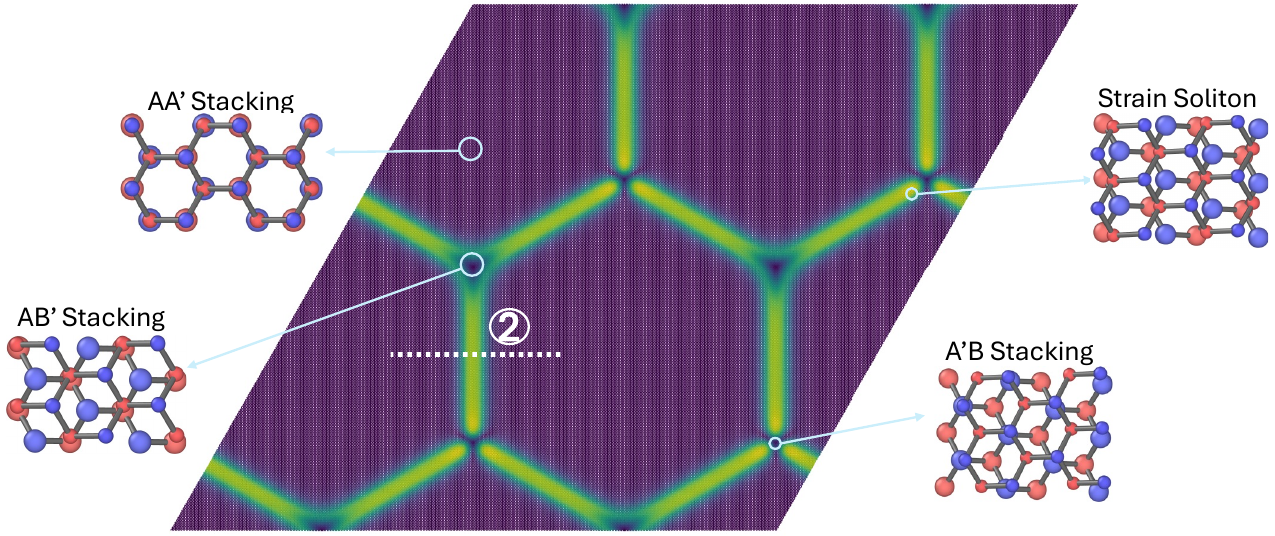}
        \label{fig:hex_net}
    }\\
    \subfloat[]
    {
        \includegraphics[width=0.5\textwidth]{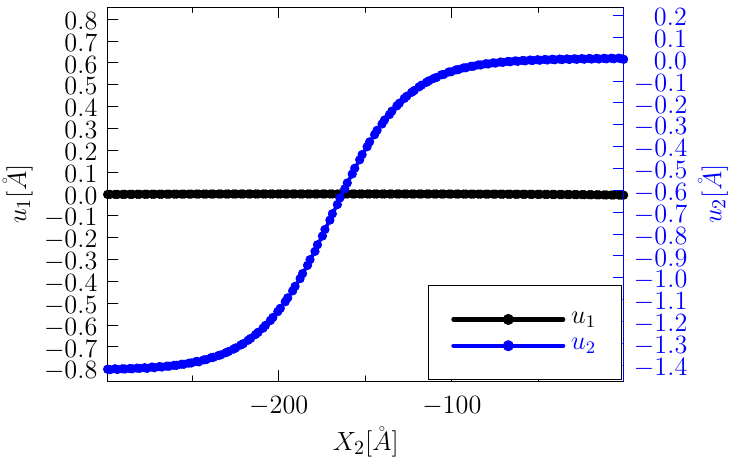}
        \label{fig:tri_scan}
    }
    \subfloat[]
    {
        \includegraphics[width=0.5\textwidth]{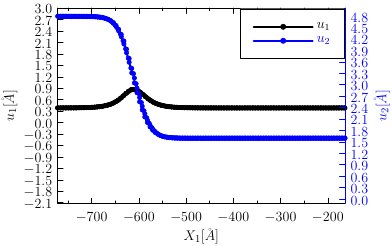}
        \label{fig:hex_scan}
    }
    \caption{Atomic strain (computed using OVITO) map of a relaxed $\SI{0.29}{\degree}$ twisted bilayer graphene \psubref{fig:tri_net} and bilayer MoS$_2$ \psubref{fig:hex_net}. Atoms are colored according to C (red), Mo (red), and S (blue). Relevant atomic stackings are included with atoms in the bottom layer depicted with larger radii. \psubref{fig:tri_scan} and \psubref{fig:hex_scan} show plots of relative displacements between the two layers measured along the dashed lines in (a) and (b).
    }
    \label{fig:bg_mos2}
\end{figure}
In this work, we focus on 2D homobilayers of graphene and MoS$_2$, two prototypical systems that exhibit distinct soliton network topologies. The structure matrices of the 2D triangular Bravais lattices of graphene and MoS$_2$ are of the form 
\begin{equation}
    \bm A = \frac{a}{2} 
    \begin{bmatrix}
        0 &  -\sqrt{3}\\
        2 & -1
    \end{bmatrix},
    \label{eq:A_matrix_inverse}
\end{equation}
where $a=\SI{2.46}{\angstrom}$ and $a=\SI{3.19702}{\angstrom}$ are the respective lattice constants. Graphene consists of two basis atoms with lattice coordinates
\begin{equation}
   \bul s^1 = \left(0,0\right), \quad  \bul s^2 = \left(\frac{1}{3},\frac{2}{3}\right),
   \label{eqn:basis_graphene}
\end{equation}
while a single layer MoS$_2$ consists of three planes of atoms formed by three basis atoms with lattice coordinates\footnote{The third lattice coordinate is described with respect to a $3\times 3$ structure matrix 
$\frac{a}{2} 
    \begin{bmatrix}
        0 &  -\sqrt{3} & 0\\
        2 & -1 & 0\\
        0 & 0 & 1
    \end{bmatrix}.$
}
\begin{equation}
   \text{S: } \bul s^1 = (0,0,0), \quad \bul s^2 = (0,0,1);\quad
   \text{Mo: } \bul s^3 = \left(\frac{1}{3},\frac{2}{3},\frac{1}{2}\right).
   \label{eqn:basis_mos2}
\end{equation}

A 2D homobilayer $\overline{\mathcal A}_{\rm t} \cup \overline{\mathcal A}_{\rm t}$ is formed by stacking a 2D lattice $\overline{\mathcal A}_{\rm t}:=\mathcal A + \{\bm s^\alpha\}$ on top of another 2D lattice $\overline{\mathcal A}_{\rm b}:=\mathcal A + \{\bm t^\alpha\}$, where $\bm s^\alpha$ and $\bm t^\alpha$ are respective basis atoms positions of the two stacked lattices.\footnote{The subscripts $\mathrm t$ and $\mathrm b$ represent 'top' and 'bottom'.}  In the case of the lowest-energy AB stacking of graphene, the shift vectors $\bm s^1$ and $\bm s^2$ of one lattice are given by \cref{eqn:basis_graphene} and those of the other graphene lattice are
\begin{equation*}
    \bul t^1 = (0,0), \quad \bul t^2 = \left(\frac{2}{3},\frac{1}{3}\right).
\end{equation*}
Similarly, in the lowest-energy AA' stacking of MoS$_2$ (local stacking of the 2H polytope), the shift vectors of the second lattice are 
\begin{equation*}
   \bul t^1 = \left(\frac{1}{3},\frac{2}{3},0 \right), \quad \bul t^2 = \left(\frac{1}{3},\frac{2}{3},1 \right), \quad
   \bul t^3 = \left(0,0,\frac{1}{2}\right).
\end{equation*}

When one of the two lattices, say $\overline{\mathcal A}_{\rm b}$, of a bilayer is subjected to an arbitrary uniform deformation with gradient $\bm F$, the heterodeformed 2D bilayer, $\overline{\mathcal A}_{\rm t}\cup \bm F \overline{\mathcal A}_{\rm b} $, forms a \emph{moir\'e Bravais lattice} given by all points where the relative displacement belongs to lattice $\mathcal A$, i.e., 
\begin{equation}
    \mathcal M = \{\bm x \in \mathbb R^2: \bm x-\bm F \bm x \in \mathcal A\}.
    \label{eqn:moire}
\end{equation}
Clearly, a moir\'e Bravais lattice exists for arbitrary heterodeformations and is independent of the basis atom positions. Moreover, the dimension of $\mathcal M$ is equal to the rank of $\bm I-\bm F$. Therefore,  if $\bm I-\bm F$ is invertible, $\mathcal M=(\bm I-\bm F)^{-1} \mathcal A$ is a 2D lattice, and if $\bm I-\bm F$ is rank one, $\mathcal M$ is a 1D lattice. On the other hand, the entire heterodeformed bilayer is periodic only for a dense collection of heterodeformations. In particular, the heterodeformed bilayer is periodic if and only if the deformation gradient is such that a 2D \emph{coincident site lattice} (CSL),
\begin{equation}
    \mathcal C := \mathcal A \cap \bm F \mathcal A,
\end{equation}
exists. In such cases, it is easy to see that $\mathcal C \subset \bm F \mathcal M$.\footnote{To see why $\mathcal C \subset \bm F \mathcal M$, let $\bm x \in \mathcal C$, which implies $\bm x \in A$ and $\bm x=\bm F \bm z$ for some $\bm z \in \mathcal A$. Since $\bm z -\bm x \in \mathcal A$, we have $\bm F^{-1} \bm x- \bm F(\bm F^{-1}\bm x) \in \mathcal A$, and it follows from definition \ref{eqn:moire} that $\bm F^{-1} \bm x \in \mathcal M$.} In other words, a smallest simulation cell of a heterodeformed bilayer is either equal to the moir\'e primitive unit cell or contains multiple moir\'e primitive unit cells, as recognized previously by Hermann \cite{Hermann_2012}.

A 2D homobilayer subjected to a small uniform heterodeformation relative to a low-energy stacking---such as the AB or BA stacking in bilayer graphene--- is \emph{not} in equilibrium. It spontaneously generates interface dislocations as a consequence of the interplay between the interface energy and the elastic energies of the two lattices. For small heterodeformations relative to the low-energy stacking, the interfacial energy increases as the relative translations between the two lattices lead to moir\'e periodic regions of low commensurability and high energy (AA stacking in bilayer graphene) interspersed with highly commensurable low-energy stackings (AB/BA in bilayer graphene). The bilayer lowers the interfacial energy through local atomic rearrangement that tends to increase the areas of high commensurability/low energy distributed periodically in the interface. As a consequence, the displacement jumps across the low-energy stackings, resulting in a periodic network of \emph{interface dislocations} or \emph{strain solitons} \cite{harley_disloc,Annevelink_2020,ahmed2024bicrystallography,ahmed2025quantifying} with translational symmetry of the moir\'e Bravais lattice. 

\Cref{fig:bg_mos2} shows strain solitons in bilayer graphene and bilayer MoS$_2$, twisted $\SI{0.29}{\degree}$ relative to their respective low energy stackings---AB in the former and AA' (local stacking of the 2H polytope) in the latter. The nodes in the soliton network correspond to high-energy stackings, while the triangular and hexagonal regions correspond to low-energy stackings. In both cases, the
Burgers vector of a dislocation, measured as a displacement jump (plotted in \Cref{fig:tri_scan,fig:hex_scan}) across the dislocation line, is parallel to the dislocation line, suggesting screw character. However, the differing topologies of the dislocation networks--- triangular in bilayer graphene and honeycomb in bilayer MoS$_2$---in \Cref{fig:tri_net,fig:hex_net} are noteworthy considering that the two systems share the same twist angle. We will now discuss how the interfacial energy, expressed as the generalized stacking fault energy (GSFE) function, determines the network topology.


\subsection{GSFE and the topology of strain soliton network}
\label{sec:gsfe_sym}
\begin{figure}[t]
    \centering
    \subfloat[]
    { 
    \includegraphics[width=0.5\textwidth]{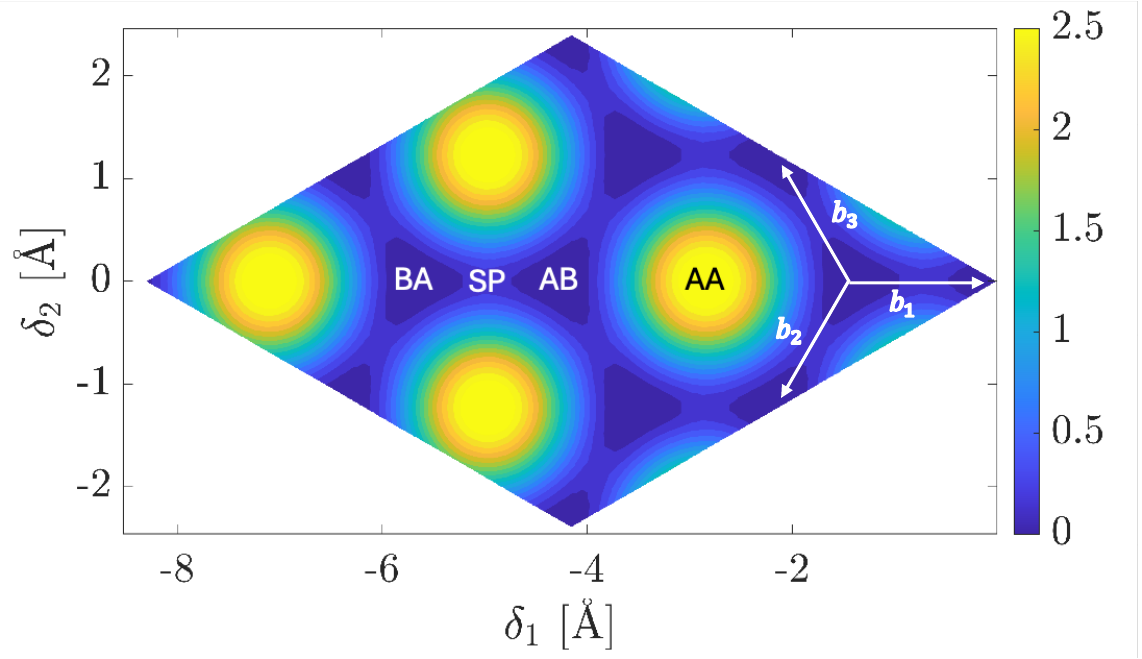}
    \label{fig:tri_GSFE}
    }
    \subfloat[]
    {
        \includegraphics[width=0.5\textwidth]{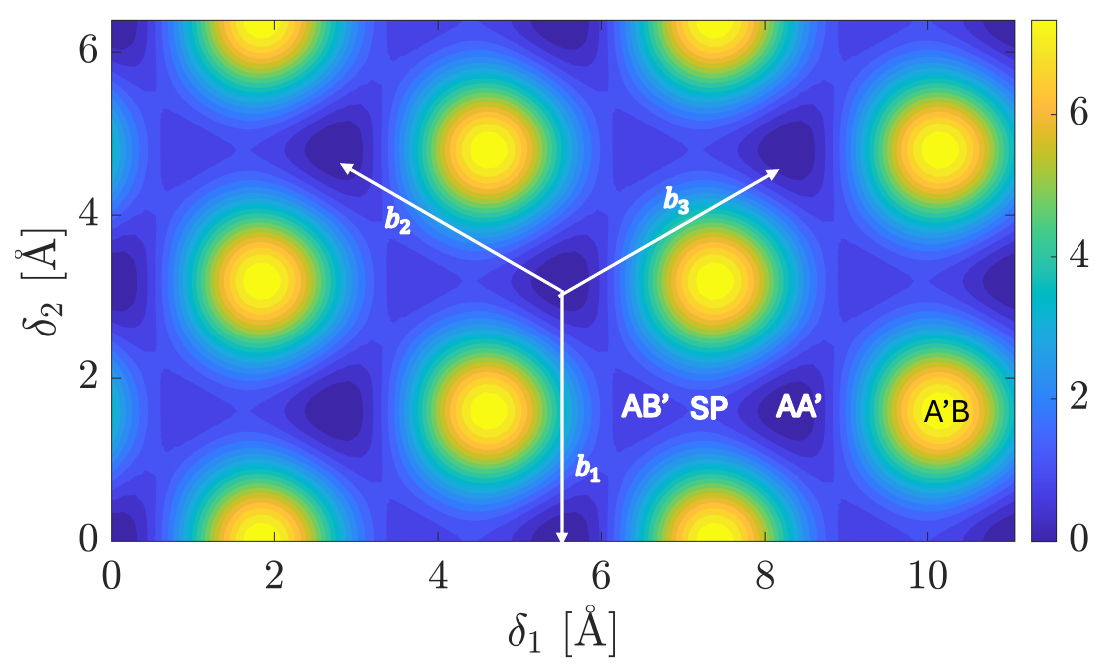}
    \label{fig:hex_GSFE}
    }
\caption{GSFEs [$\si{\milli\electronvolt\per\angstrom\squared}$] of bilayer graphene and bilayer MoS$_2$. $\delta_1$ and $\delta_2$ represent relative translations between the two layers. The local minima represent low-energy stackings---AB/BA in bilayer graphene and AB'/BA' in bilayer MoS$_2$.}
\label{fig:tri_vs_hex_dis}
\end{figure}

The interfacial energy is given by the GSFE, a periodic function of the relative in-plane displacements between the constituent 2D materials, measured before the heterodeformation is applied. \Cref{fig:tri_vs_hex_dis} shows plots of the GSFE for bilayer graphene and MoS$_2$, wherein the energy scale is shifted such that the zero is the energy minimum. 

A uniform heterodeformation $\bm F$ imposed on one of the layers induces a spatially varying relative displacement field $\bm u(\bm x)$ between the two layers. Writing the displacement field before relaxation as $\bm u(\bm x) = (\bm F - \bm I)\bm x$, the image of this map determines which relative translations are sampled. When $\bm F - \bm I$ has rank 2,\footnote{
    This includes heterodeformations containing a rotational component (obtained from the polar decomposition $\bm F=\bm R\bm U$ with $\bm R\neq\bm I$), but also more general anisotropic stretches for which $\bm F-\bm I$ remains full rank.
} the relative translation field samples the full unit cell of the GSFE domain. In such cases, regions corresponding to neighborhoods of GSFE minima expand into low-energy domains, while narrow transition regions interpolate between adjacent minima. These transition regions form a periodic 2D soliton network.

Comparing \Cref{fig:bg_mos2,fig:tri_vs_hex_dis}, we note that the \emph{topology} (number of sides of the polygon) of a 2D soliton network is entirely determined by the number of nearest-neighbor minima surrounding a local minimum---three in \Cref{fig:tri_GSFE}, and six in \Cref{fig:hex_GSFE}. Alternatively, the network topology can also be deduced by focusing on the GSFE's local maxima, whose corresponding high-energy stackings occur at the nodes of the 2D soliton network. For example, each local maximum in \Cref{fig:tri_GSFE} is surrounded by six local minima, suggesting that a node in a relaxed bilayer graphene is surrounded by six low-energy stackings. Similarly, three local minima neighboring a local maximum in \Cref{fig:hex_GSFE} justify the hexagonal network in MoS$_2$. Interestingly, not all local maxima in \Cref{fig:hex_GSFE} are equivalent as the energy of the A'B stacking is higher than that of the AB' stacking. This asymmetry is evident in \Cref{fig:hex_net}, where there is a noticeable difference between the nodes corresponding to the A'B and AB' stackings.

The dimension of the soliton network is that of the moir\'e Bravais lattice, which is equal to the rank of $\bm F-\bm I$. If $\bm F-\bm I$ has rank 1, the relative displacement field varies only along a single direction in real space, and the bilayer samples only a one-dimensional path in the GSFE domain rather than the full 2D displacement space. In such cases, the relaxed configuration consists of a periodic array of parallel soliton lines, forming a 1D network instead of a polygonal 2D network. 

Summarizing, the rank of $\bm F-\bm I$ determines the network dimensions. The topology of a 2D soliton network is determined by  the point group symmetry of the GSFE minima. In addition, the topology is invariant under small changes in the heterodeformation: triangles remain triangles and honeycombs remain honeycombs, although their geometric shapes and sizes may vary continuously with $\bm F$.

\subsection{The Burgers vectors of 1D and 2D strain soliton networks}
\label{sec:geom_des}
Each dislocation in a strain soliton network is described by a Burgers vector-line vector pair, $(\bm b,\bm l)$. While the dislocation lines are heterodeformation-dependent, their Burgers vectors are determined entirely by the GSFE. In particular, the vectors connecting any two nearest GSFE minima form the set of admissible Burgers vectors. Therefore, we have 
\begin{subequations}
\label{eqn:b_admissible}
\begin{align}
    \text{bilayer graphene: }
    \bul b^i &\in 
        \left\{ \left (-\frac{1}{3},-\frac{2}{3} \right), \left (-\frac{1}{3},\frac{1}{3}\right), \left(\frac{2}{3},\frac{1}{3}\right)
        \right\}, \label{eqn:b_graphene}\\
    \text{bilayer MoS$_2$: }
    \bul b^i &\in 
        \left\{(-1,0), (1,1), (0,-1)\right\}. \label{eqn:b_mos2}
\end{align}
\end{subequations}
which are shown in \Cref{fig:tri_vs_hex_dis}. Since the Burgers vectors in bilayer graphene have fractional lattice coordinates, they are \emph{not} in $\mathcal A$, and therefore, the soliton network consists of partial dislocations. On the other hand, dislocations in bilayer MoS$_2$ are full dislocations. 

\begin{figure}[t]
    \centering
    \subfloat[]{
        \includegraphics[width=0.5\textwidth]{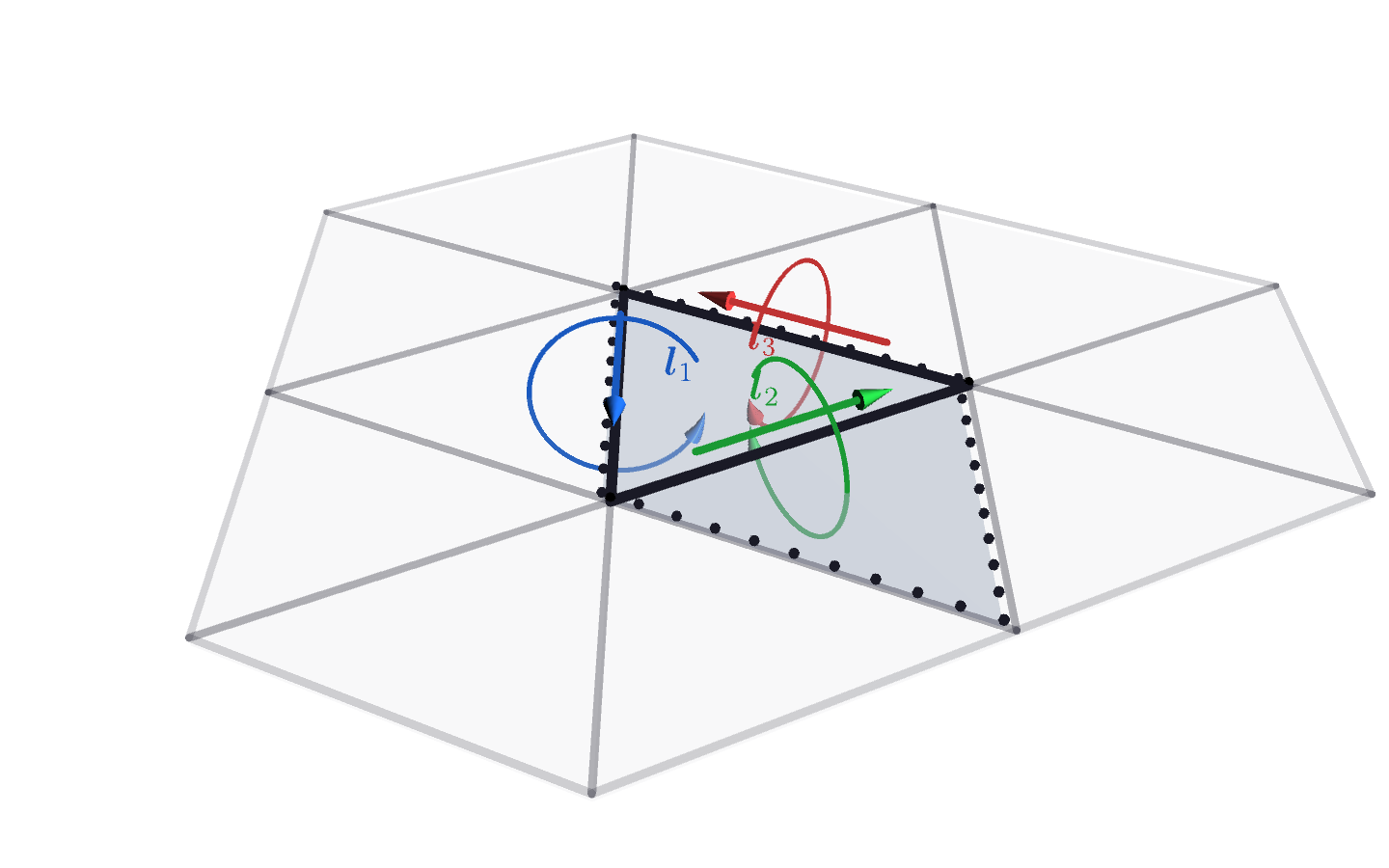}
        \label{fig:tri_circuit}
    }
    \subfloat[]{
        \includegraphics[width=0.5\textwidth]{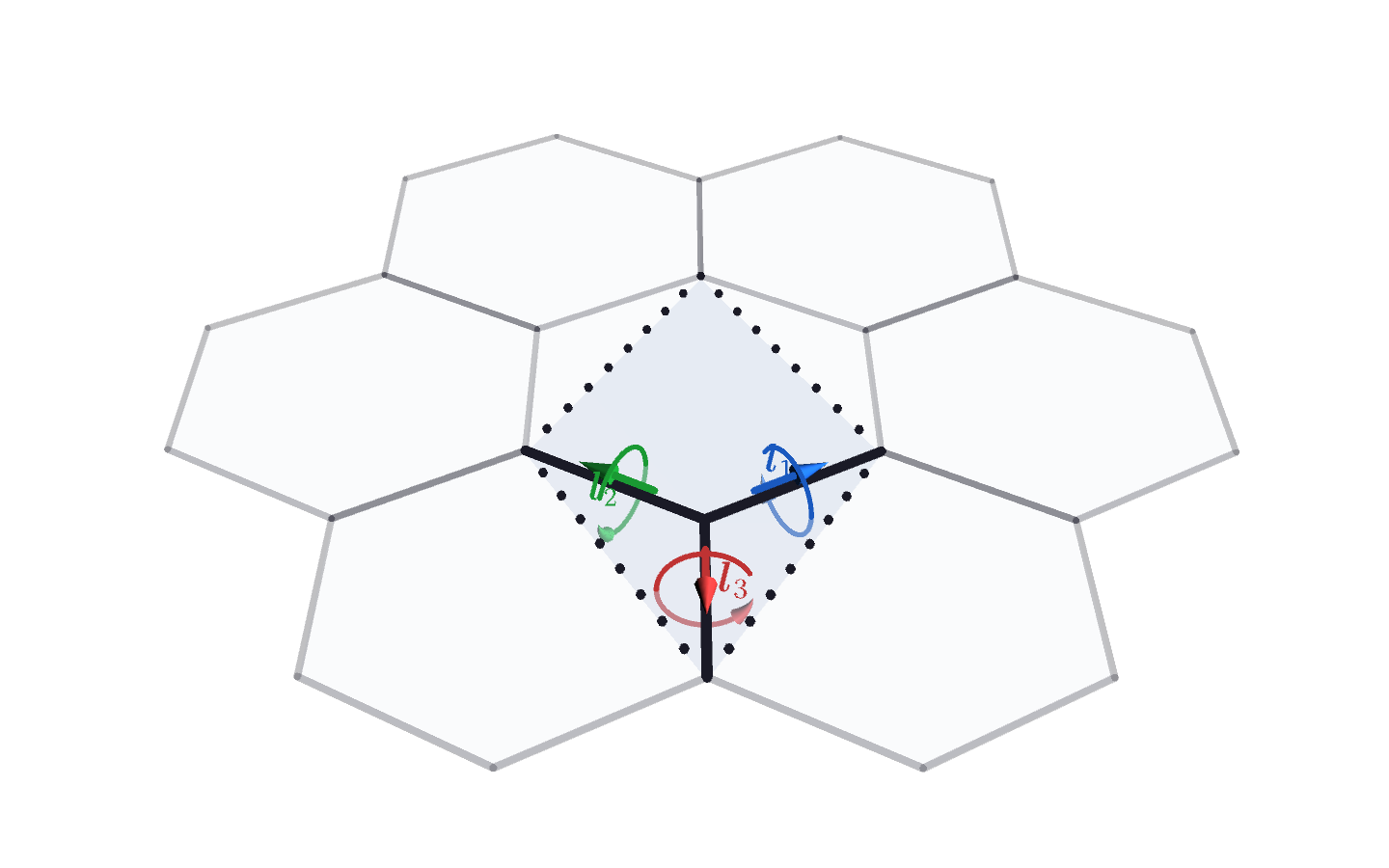}
        \label{fig:hex_circuit}
    }
    \caption{Strain soliton networks in bilayer graphene and MoS$_2$ consist of moir\'e-periodic collection of three dislocation lines (solid black) in the primitive unit cell (shaded) of the moir\'e Bravais lattice of the bilayer. The Burgers circuit analysis of the three dislocations with depicte line directions yields $\bm b^1+\bm b^2+\bm b^3=\bm 0$.}
    \label{fig:circuit}
\end{figure}
The 2D soliton networks in \Cref{fig:circuit} are moir\'e periodic, and can be described by three distinct line vectors---$\bm l^1$, $\bm l^2$, and $\bm l^3$---in the primitive unit cell (shaded cell) of the moir\'e Bravais lattice, as shown in \Cref{fig:circuit}. In the triangular soliton network of bilayer graphene, the three lines themselves are any set of primitive lattice vectors of the moir\'e Bravais lattice, while in bilayer MoS$_2$, their differences form a collection of three primitive vectors of the moir\'e Bravais lattice. Due to the freedom associated with the equivalence $(\bm b,\bm l) \sim (-\bm b,-\bm l)$, we choose line directions as depicted in \Cref{fig:circuit}, which satisfy the condition
\begin{equation}
    \sum_{i=1}^3 \bm l^i = \bm 0.
    \label{eqn:lsum}
\end{equation}
Since the line directions in \Cref{fig:tri_net} are the sides of a triangle, they clearly satisfy \cref{eqn:lsum}. However, the line directions in a hexagonal network need not satisfy \cref{eqn:lsum} as the triple junction of the lines in \Cref{fig:hex_net} can lie anywhwere in the shaded moir\'e primitive unit cell. Nevertheless, we will assume \cref{eqn:lsum} for both networks to  ensure our inverse design problem, discussed in the next section, is well-posed. A Burgers circuit analysis with the above choice of line directions gives us the condition 
\begin{equation}
    \sum_{i=1}^3 \bm{b}^i= \bm 0.
    \label{eqn:bsum}
\end{equation}

\begin{figure}[H]
    \centering
    \subfloat[]{
        \includegraphics[height=0.2\textwidth]{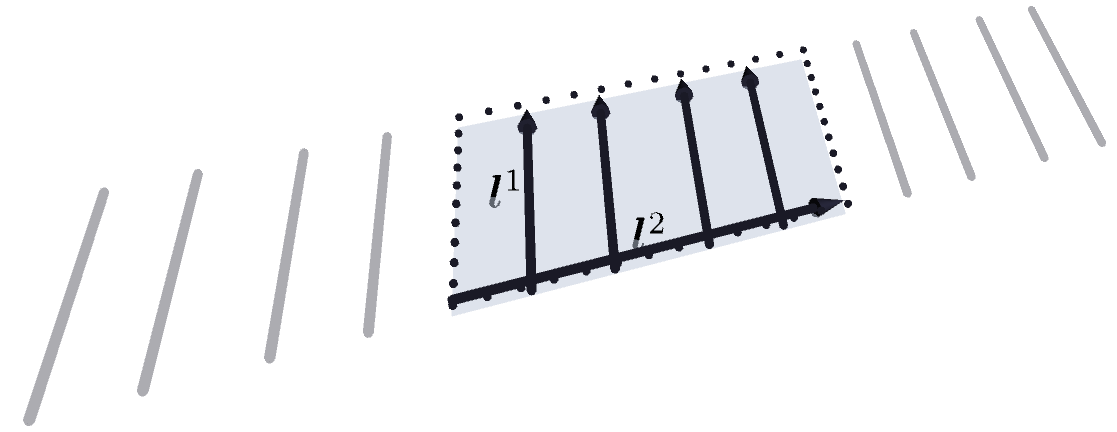}
        \label{fig:1d_circuit}
    }
        \subfloat[]
    {
        \includegraphics[height=0.25\textwidth]{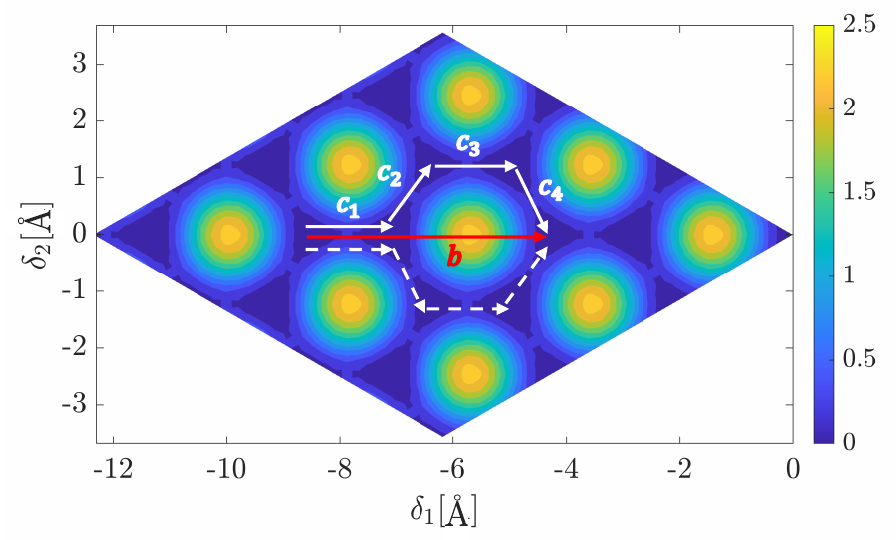}
        \label{fig:GSFE_1D}
    }
    \caption{\psubref{fig:1d_circuit} A moir\'e-periodic 1D strain soliton network in bilayer graphene showing four parallel dislocation lines in the primitive moir\'e unit cell (shaded region). The sum of the Burgers vectors, $\bm b^1=\bm c^1+\bm c^2+\bm c^3+\bm c^4$, of the lines is equal to a lattice vector in $\mathcal A$. \psubref{fig:GSFE_1D} The four Burgers vectors in the unit cell correspond to a GSFE path that connects two energy wells separated by $\bm b$.
    }
    \label{fig:1D_strain_soliton_origin}
\end{figure}

While a 2D strain soliton network is described as a collection of three dislocation lines within a primitive unit cell of a moir\'e Bravais lattice, a 1D soliton network includes a sequence of, say $n$, parallel dislocation lines, as shown in \Cref{fig:1d_circuit}. Therefore, we take $\bm b^2=\bm b^3=\bm 0$. The $n$ dislocations are parallel to $\bm l^1$, and the moir\'e Bravais lattice is spanned by $\bm l^1$ and $\bm l^2$. Since the dislocation lines are parallel, their Burgers vectors---say $\bm c^1,\dots,\bm c^n$---can be summed up to give $\bm b^1=\bm c^1+\dots+\bm c^n$. Periodic boundary conditions impose the restriction that $\bm b^1$ has to be equal to a lattice vector of $\mathcal A$. 

The order of the Burgers vectors of the $n$ lines corresponds to a path in the GSFE that connects two energy wells separated by the total Burgers vector $\bm b^1$. For example, \Cref{fig:1d_circuit} shows four dislocations in a moir\'e unit cell. The red arrow in the GSFE of bilayer graphene, shown in \Cref{fig:GSFE_1D}, depicts the total Burgers vector, which splits into $n=4$ dislocation lines shown in \Cref{fig:1d_circuit}. The heterodeformation is determined by the total Burgers vector $\bm b^1$ and the lines $\bm l^1$ and $\bm l^2$, and \emph{not} on how $\bm b^1$ decomposes. In other words, different decompositions of $\bm b^1$ correspond to the same heterodeformation. For instance, an energetically equivalent alternate path (dotted), shown in \Cref{fig:GSFE_1D}, yields an alternate arrangement of dislocations. Therefore, we will use the pairs $(\bm b^1,\bm l^1)$ and $(\bm 0, \bm l^2)$ to describe a 1D soliton network. This ensures that the inverse-design problem, discussed in the next section, is well-posed.

\section{From soliton network to heterodeformation}
\label{sec:int_algebra}
In the previous section, we noted that a soliton network is described by a periodic arrangement of dislocation lines with the translational symmetry of the moir\'e Bravais lattice. In this section, we pose the following inverse-design problem:
\begin{quote}
    \emph{For a given soliton network, identify the heterodeformation that produces this network as the system undergoes structural relaxation.}
\end{quote}
To solve the above problem we have to ensure that the problem is well-posed, i.e., a one-to-one mapping exists between soliton network descriptions and heterodeformations. Without loss of generality, we will assume a heterodeformation is applied by imposing a uniform deformation to the bottom lattice while the top lattice remains fixed, i.e., $\overline{\mathcal A}_{\rm t} \cup \bm F \overline{\mathcal A}_{\rm b}$ is an unrelaxed heterodeformed bilayer.\footnote{While the top and bottom lattices can be deformed independently (using deformation gradients $\bm F_{\rm t}$ and $\bm F_{\rm b}$), the soliton geometry up to a uniform deformation is determined by the relative deformation $\bm F_{\rm b}^{-1} \bm F_{\rm t}$.} We

Solving for the heterodeformation alone is insufficient if we ultimately want to simulate the structural reconstruction of a bilayer under periodic boundary conditions (PBCs), which can be enforced if and only if a 2D CSL exists, i.e., $\mathcal A \cap \bm F \mathcal A$ is a 2D lattice. The existence of a 2D CSL is equivalent to the condition \cite{Hermann_2012,koda2016coincidence,ADMAL2022} that the \emph{transition matrix}
\begin{equation}
    \bm Q:= \bm A^{-1}\bm F^{-1} \bm A
    \label{eqn:rational}
\end{equation}
is a rational matrix. Therefore, we will focus our attention on heterodeformation gradients $\bm F$ that satisfy \cref{eqn:rational}. Moreover, we need the CSL primitive vectors, $\bm \ell^1$ and $\bm \ell^2$, that span the smallest simulation domain on which PBCs are enforced. In an earlier work \cite{ADMAL2022}, we showed that the CSL primitive vectors (and various other geometric properties) follow from the transition matrix using Smith normal form (SNF) bicrystallography \cite{ADMAL2022}. SNF bicrystallography utilizes the Smith normal form for integer matrices to generalize the notions of least common multiple (lcm) and greatest common divisor (gcd) of arbitrary lattices. In particular, the CSL is obtained as an lcm lattice of the lattices $\mathcal A$ and $\bm F\mathcal A$. Therefore, our ultimate goal in this section is to start with a network geometry and inverse-design the rational transition matrix to construct the heterodeformed bilayer with PBCs.

\subsection{Inverse design of heterodeformations}
\label{sec:def_top_inv}
Starting with a desired strain soliton network, described by three Burgers vector-line vector pairs, 
\begin{equation}
    \{(\bm b^i,\bm l^i\}:i=1,2,3\},
    \label{eqn:bl_pairs}
\end{equation}
we now seek a unique heterodeformation that yields this network. If the network is 2D, recall that the Burgers vectors in \cref{eqn:bl_pairs} are unique and belong to the admissible set in \cref{eqn:b_admissible}, and the line vectors should satisfy \cref{eqn:lsum}. On the other hand, if the network is 1D, $\bm b^1$ is any lattice vector of the Bravais lattice $\mathcal A$, and $\bm b^2=\bm b^3 = \bm 0$. We will always assume the line vectors have rational lattice coordinates with respect to $\mathcal A$. We will later show that this assumption is necessary and sufficient for \cref{eqn:rational} to hold. Under this setting, the dislocation density is given by the Nye tensor \citep{nye1953some} $\bm \alpha$: 
\begin{equation}
    \bm \alpha^{\rm T}
    =\delta_{x_3}\frac{1}{a} \sum_{i=1}^3 \bm b^i \otimes \bm l^i,
    \label{eqn:nye_burgers_line}
\end{equation}
where $\delta_{x_3}$ is the 1D Dirac delta distribution along $x_3$, and $a$ is the area of the primitive unit cell of the moir\'e Bravais lattice, given by
\begin{align}
    \text{1D network: } a= |\bm l^1 \times \bm l^2|; \quad
    \text{2D network: } a = \begin{cases}
        |\bm l^1 \times \bm l^2| & \text{ bilayer graphene,} \\
        3 |\bm l^1 \times \bm l^2| & \text{ bilayer MoS$_2$}.
    \end{cases}
    \label{eqn:area_calc}
\end{align}

From \cref{eqn:nye_burgers_line}, it is clear that the conditions in \cref{eqn:bsum,eqn:lsum} on the Burgers vectors and line vectors ensure that vector pairs in \cref{eqn:bl_pairs} uniquely define $\bm \alpha$. This one-to-one correspondence will ensure that the inverse design problem is well-posed considering that we will now express $\bm \alpha$ in terms of the deformation gradient $\bm F$.

Since $\bm \alpha = \curl(\InvFe)$, where $\bm F$ is expressed as a discontinuous function
\begin{equation}
    \bm F(\bm x) = \begin{cases}
        \bm I & \text{ if $x_3>0$}, \\
        \bm F &\text{ if $x_3 < 0$},
    \end{cases}
    \label{eqn:piecewiseF}
\end{equation}
we have
\begin{equation}
    \bm \alpha = \delta_{x_3} 
    \begin{bmatrix}
        -\jump{\InvFe_{12}}  &  -\jump{\InvFe_{22}} \\
        \jump{\InvFe_{11}}  &  \jump{\InvFe_{21}}
    \end{bmatrix}.
    \label{eq:twist_nye_ten_in}
\end{equation} 
Here, $\jump{\InvFe_{km}}$ denotes the jump in the $km$-component of $\InvFe$. Taking, the transpose of \cref{eq:twist_nye_ten_in} and by further manipulation, we obtain 
\begin{align}
    \bm \alpha^{\rm T}
    &= \delta_{x_3}  
    \begin{bmatrix}
        \jump{\InvFe_{11}}  &  \jump{\InvFe_{12}} \\
        \jump{\InvFe_{21}}  &  \jump{\InvFe_{22}} 
    \end{bmatrix}
    \begin{bmatrix}
        0 & 1 \\
        -1 & 0
    \end{bmatrix} \notag \\ 
    &=\delta_{x_3} (\bm I-\InvFe)\bm R_{-\pi/2},
    \label{eq:nye_def}
\end{align}
where $\bm R_{-\pi/2}$ denotes 2D rotation by $-\pi/2$ radians. 

Comparing \cref{eqn:nye_burgers_line,eq:nye_def}, we obtain the following relationship between the heterodeformation and the strain soliton network
\begin{align}
    (\bm I-\InvFe) \bm R_{-\pi/2}=\frac{1}{a}\sum_{i=1}^3 \bm b^i \otimes \bm l^i.
    \label{eq:def_burgers_id}
\end{align}
While \cref{eq:def_burgers_id} yields the desired $\bm F$, and from it the transition matrix follows using \cref{eqn:rational}, this sequence of steps to obtain $\bm Q$ is fraught with round-off errors. In Section S-1 of Supplementary Information, we show that the transition matrix can be obtained directly from the lattice coordinates of the vector pairs in \cref{eqn:bl_pairs}:
\begin{equation}
    \bm Q = \bm I-\sum_{i=1}^3 \frac{\bul b^i \otimes \bol m^i}{\beta z^i},
    \label{eq:Q_final_cal}
\end{equation}
where $\bm m^i \in \mathcal A_{\rm t}^*$ is the smallest reciprocal vector orthogonal to $\bm l^i$  with coordinates 
\begin{subequations}
\label{eqn:mz}
\begin{align}
    \bol m^i&=(\bul l^i_2, -\bul l^i_1)/\gcd(\underline l^i_1,\underline l^i_2), \text{ and} \\
    z^1 := |\bul l^2 \cdot \bol m^1|, \quad
    z^2 &:= |\bul l^3 \cdot \bol m^2|,\quad 
    z^3 := |\bul l^1 \cdot \bol m^3|
\end{align}
\end{subequations} 
are rational numbers.\footnote{
    The gcd of two rational numbers $\frac{a}{b}$ and $\frac{c}{d}$ is $\frac{\gcd(ad,cb)}{bd}$.
} 
The constant $\beta=a/|\bm l^1 \times \bm l^2|= 1$ or $3$, depending on the cases described in \cref{eqn:area_calc}. The heterodeformation gradient then follows from \cref{eqn:rational,eq:Q_final_cal}. From \cref{eq:Q_final_cal,eqn:mz}, it is clear that choosing line vectors with rational integer coordinates ensures that the transitino matrix is rational.

Finally, we obtain the CSL primitive vectors (simulation box vectors) $\bm \ell^1$ and $\bm \ell^2$ from the rational transition matrix $\bm Q$ by employing SNF bicrystallography. 
The following theorem summarizes the results of this section.

\begin{theorem}
    A uniformly heterodeformed bilayer $\overline{\mathcal A} \cup \bm F \overline{\mathcal A}$ hosts a strain soliton network $\{\bm b^i,\bm l^i:i=1,2,3\}$, where $\bm b^i$ are bilayer's admissible Burgers vectors and $\bm l^i$ are finite dislocation lines with rational lattice coordinates (with respect to $\mathcal A)$ if and only if 
    \begin{equation}
        \Fe = \bm A \bm Q^{-1} \bm A^{-1}, \text{ where }
        \bm Q = \bm I-\sum_{i=1}^3 \frac{\bul b^i \otimes \bol m^i}{\beta z^i},
    \label{eqn:fin_def}
    \end{equation}
    $\bm m^i$ is reciprocal lattice vector orthogonal to $\bm l^i$ and $z^i$ are rationals, given in \cref{eqn:mz}. $\beta=1$ if the network is 1D or triangular, and $\beta=3$ for hexagonal networks.
\end{theorem}

It is interesting to note that if the rational lattice coordinates of $\bm l^i$ are integers (equivalently, if $\bm l^i \in \mathcal A$), then from \cref{eq:Q_final_cal,eqn:mz} it follows that $\bm Q \bul l^i$ contains integers. From the definition of $\bm Q$, this implies $\bm l^i \in \bm F \mathcal A$. In other words, $\bm l^i$ is a CSL vector. Therefore, we conclude that if the line vectors have integer coordinates, then they can be chosen as the simulation box vectors. In such cases, we refer to the resulting soliton network as a \emph{simple network}. On the other hand, if at least one of the line vectors has non-integer rational coordinates, then the smallest simulation box necessarily contains multiple moir\'e unit cells, and we refer to the resulting network as \emph{complex network}.

Algorithm~\ref{alg:inverse_design} describes the entire inverse design framework---starting from a target strain soliton network, it constructs a heterodeformation and the box vectors of the structural relaxation simulation cell, which yields a bilayer with the desired strain soliton network.
\begin{algorithm}[H]
\caption{Given a lattice $\mathcal A$ and a target strain soliton network, compute the heterodeformation and the periodic simulation cell of the heterodeformed bilayer.}
\label{alg:lattice_transformation}
\begin{algorithmic}[1]

\State \textbf{Input:} Structure matrix $\bm A$ and the soliton network $\{(\bul b^i,\bul l^i): i=1,2,3\}$.

\State Compute the variables $a$, $\bm m^i$, and $z^i$ using \cref{eqn:area_calc,eqn:mz}.

\State Compute the transition matrix $\bm Q$ using \cref{eq:Q_final_cal}

\State Compute the heterodeformation gradient $\bm F$ using \cref{eqn:fin_def}.

\If{each $\bul l^i$ has integer lattice coordinates}
    \State Choose $\bm \ell^1= \bm A \bul l^1$ and $\bm \ell^2 = \bm A \bul l^2$ as the periodic simulation cell vectors.
\Else
    \State Compute the primitive simulation cell vectors $\bm \ell^1$ and $\bm \ell^2$ using SNF bicrystallography \cite{ADMAL2022}\footnotemark.
\EndIf

\State \textbf{Output:} Heterodeformation gradient $\bm F$, and simulation cell primitive vectors $\bm \ell^1$ and $\bm \ell^2$.
\label{alg:inverse_design}
\end{algorithmic}
\end{algorithm}
\footnotetext{SNF bicrystallography is a robust framework that leverages integer matrix algebra to automate the generation of crystal interfaces (grain boundaries and 2D materials interfaces) and enumerate interface dislocations. This approach yields dimension-independent algorithms applicable to any crystal system. The framework is implemented in the open-source C++ library \texttt{oILAB}, which is accompanied by Python bindings and is available at \url{https://github.com/admal-research-group/oILAB.git}.}

\section{Construction of targeted 1D and 2D strain soliton networks}
\label{sec:non_trivial}
In this section, we demonstrate our inverse-design framework to construct simple and complex strain soliton networks in bilayer graphene and bilayer MoS$_2$. As outlined in Algorithm~\ref{alg:inverse_design}, the target network is the input, and the output is a uniformly heterodeformed bilayer that, upon structural relaxation, yields the desired network. The lattice vector pairs of the input network (\cref{eqn:bl_pairs}) are described in terms of their rational lattice coordinates with respect to the lattices introduced in \Cref{sec:geom_str}. 

To confirm our inverse design framework, we use atomistic simulations to minimize the uniformly heterodeformed bilayers. Simulations are performed using Large-Scale Atomic/Molecular Massively Parallel Simulator (LAMMPS) \citep{thompson2022lammps}. Periodic boundary conditions are imposed along the $X_1$ and $X_2$ directions, and shrink-wrapped boundary conditions in the out-of-plane direction. For bilayer graphene, the reactive empirical bond order (REBO) potential \citep{Rebo_Brenner_2002} is used to model the intralayer bonding in each graphene layer while the registry-dependent Kolmogorov--Crespi (KC) potential \citep{ouyang2018nanoserpents} describes the interlayer van der Waals (vdW) interaction. Table~\ref{table:parameters} lists the parameters for the KC potential.
\begin{table}[t]
\centering
\small
\begin{tabular}{c c c c}
\hline
Parameter & Value & Parameter & Value \\
\hline
$C$ & $\SI{6.678908e-4}{\meV}$ & $A$ & $\SI{12.660270}{\meV}$ \\
$C_0$ & $\SI{21.847167}{\meV}$ & $\delta$ & $\SI{0.771810}{\angstrom}$ \\
$C_2$ & $\SI{12.060173}{\meV}$ & $\lambda$ & $\SI{3.143921}{\per\angstrom}$ \\
$C_4$ & $\SI{4.711099}{\meV}$ & $z_0$ & $\SI{3.328819}{\angstrom}$ \\
\hline
\end{tabular}
\caption{Parameters of the KC potential.}
\label{table:parameters}
\end{table}
For bilayer MoS$_2$, the intralayer nearest-neighbor interaction between molybdenum and sulfur atoms is modeled using the modified Stillinger--Weber (sw/mod) potential \citep{jiang2015parametrization}. The interlayer van der Waals interaction is modeled using the registry-dependent interlayer potential (ILP) \cite{ouyang2021anisotropic} with a cutoff radius of $16~\si{\angstrom}$ to ensure adequate sampling of dispersive interactions. As the structural reconstruction was conducted without an external electric field, explicit Coulomb interactions were intentionally omitted. Atomic reconstruction is simulated by minimizing the total energy using the fast inertial relaxation engine (FIRE) algorithm \citep{bitzek2006fire} with an energy tolerance of $\SI{1e-20}{\eV}$ and a force tolerance of $\SI{1e-20}{\eV \per \angstrom}$.

All relaxed atomistic configurations of heterodeformed bilayers in this section are visualized as color density plots with colors representing the lattice strain norm, $\sqrt{\epsilon_{11}^2+\epsilon_{22}^2+2\epsilon_{12}^2}$. The lattice strain tensor is measured relative to the uniformly heterodeformed bilayer and computed using OVITO.

\subsection{\label{sec:trivial_moire}1D strain soliton networks}
We begin with the simplest example of a 1D soliton in MoS$_2$ consisting of a single dislocation with $\bul b^1 = (0,1)$  and $\bul l^1 = (-1,1)$.  Recall that for 1D networks, we always choose $\bm b^2=\bm b^3=\bm 0$. The corresponding line vectors are chosen as $\bul l^{2}= (121,119)$ and $\bul l^3=(-120,-120)$, leading the to transition matrix
\begin{equation}
    \bm Q = \frac{1}{240}
    \begin{bmatrix}
        240 & 0 \\
        -1 & 239 
    \end{bmatrix},
\end{equation}
and the deformation gradient is 
\begin{equation}
    \bm F=\begin{bmatrix}
        1.006276 & -0.003624 \\
        0.003624 & 0.997908
        \label{eq:1D_moire_MOS2}
    \end{bmatrix}
    = \bm R(0.207179^\circ) 
    \begin{bmatrix}
        1.006283 & -0.000015 \\
        -0.000015 & 0.997915
    \end{bmatrix},
\end{equation}
where the latter equation is the polar decomposition of $\bm F$ into a twist and a stretch tensor.

\begin{figure}[t]
    \centering
    \subfloat[]
    {
    \includegraphics[height=0.5\linewidth]{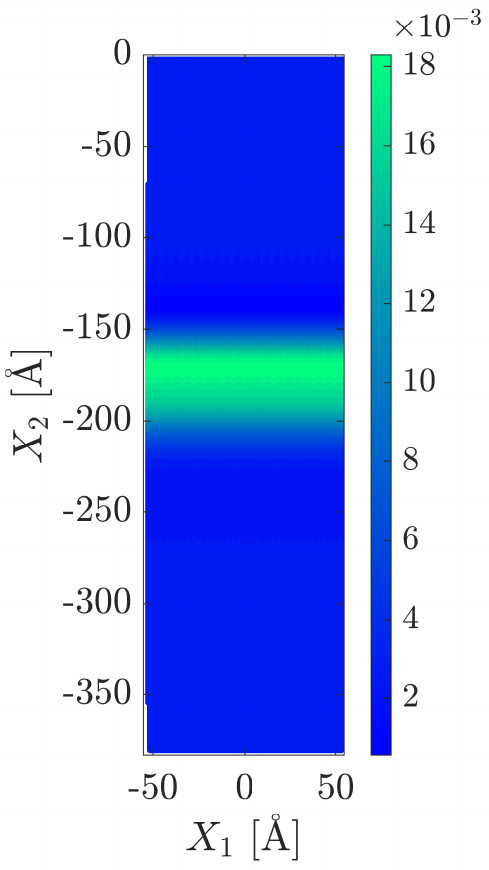}
    \label{fig:shear_MOS2}
    }
    \subfloat[]
    {
        \includegraphics[height=0.5\linewidth]{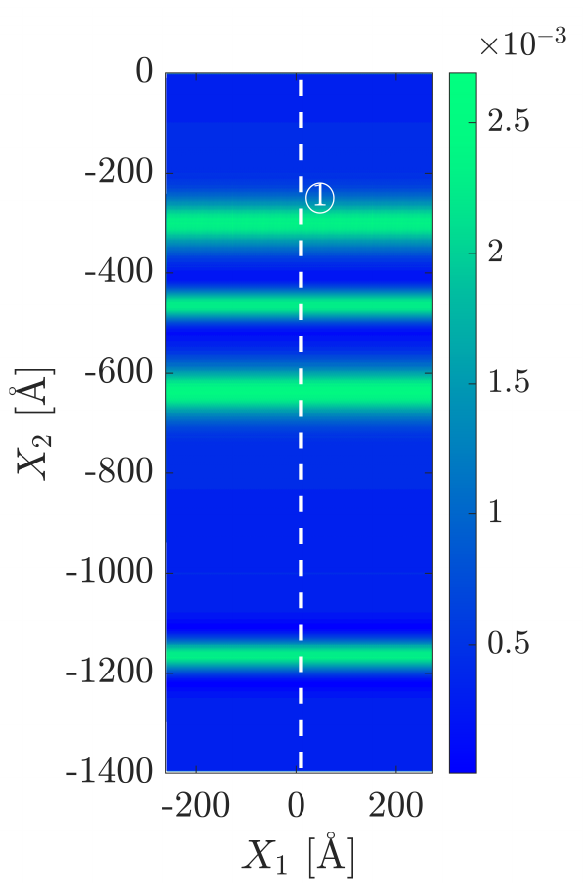}
        \label{fig:simple_shear}
    }
    \subfloat[]
    {
    \includegraphics[height=0.53\linewidth]{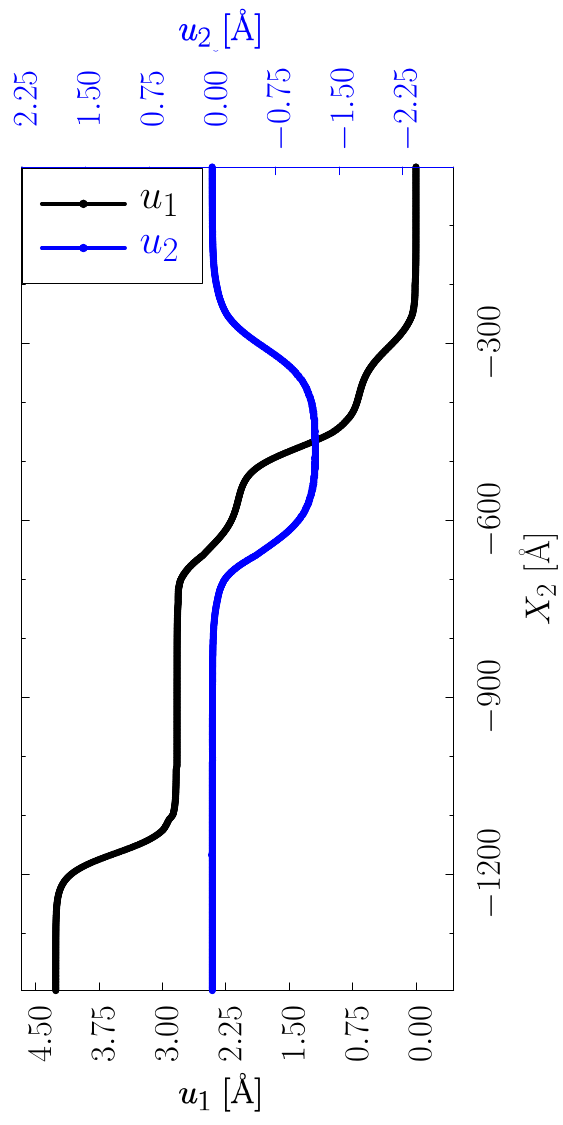}
    \label{fig:simple_burgers}
    }
    \caption{1D strain solitons in \psubref{fig:shear_MOS2} bilayer MoS$_2$ and  \psubref{fig:simple_shear} bilayer graphene with total Burgers vector $\bul b^1=(1,0)$ (mixed dislocation) and $\bm b^2=(-1,1)$ (screw dislocation), respectively. While the former occurs as a single mixed dislocation, the screw in the latter splits into two screw partials and two mixed partials. \protect\subref{fig:simple_burgers} shows plots of the displacement components, measured along line \textcircled{1} in \protect\subref{fig:simple_shear}. The displacements are measured relative to the AB stacked bilayer graphene---the stacking prior to the heterodeformation.}
    \label{fig:disloc_1D}
\end{figure}

Since there exists no lattice vector smaller than the lattice vector $\bm b^1$, we expect the dislocation to be stable as it cannot split. \Cref{fig:shear_MOS2} shows the relaxed configuration. It contains a single mixed dislocation and, as expected, is stable. 

Next, we examine 1D solitons in bilayer graphene by choosing $\bm b^1=(-1,1)$, $\bm l^1=(-1,1)$, and $\bm l^2=(569,567)$. The resulting transition matrix and deformation gradient are
\begin{equation}
    \bm Q = \frac{1}{1136}
    \begin{bmatrix}
        1137 & 1 \\
        -1 & 1135 
    \end{bmatrix},
\end{equation}
and
\begin{equation}
    \Fe_{\rm t}=\begin{bmatrix}
        1.001320 & -0.000762 \\
        0.002287 & 0.998680
    \end{bmatrix}= 
    \bm R(0.087358^\circ) 
    \begin{bmatrix}
        1.001323 & 0.000760 \\
        0.000760 & 0.998680
\end{bmatrix}.
        \label{eq:1D_moire_BG}
\end{equation}
The relaxed configuration of the heterodeformed bilayer graphene, shown in \Cref{fig:simple_shear}, clearly shows the decomposition of the dislocation into two mixed partials and two screw partials. The underlying mechanism of this decomposition can be understood from the path shown in the GSFE plot in \Cref{fig:GSFE_1D}. \Cref{fig:simple_burgers} further confirms this by showing the displacement components along the scanning direction \textcircled{1}, measured with respect to the AB stacked bilayer. 

It is important to note that there are two competing energies at play here---dislocation core energies and interaction energies. Since the latter increases as the average distance between dislocations decreases, it is energetically favorable for the dislocation to \emph{not} decompose if the dimension along $\bm l^2$ is decreased, leading to strong interactions with periodic images. For instance, in a bilayer graphene with $\bul l^{2}= (43,41)$, we did not observe dislocation decomposition.

\subsection{\label{sec:twist_moire}2D strain soliton networks}
In this section, we present case studies of heterodeformations that lead to simple and complex 2D soliton networks in bilayer graphene and bilayer MoS$_2$. In particular, we will demonstrate the construction of heterodeformations shown in \Cref{fig:disloc_degenracy_unrel}, which share a moir\'e Bravais lattice but have unique soliton networks.

\subsubsection{\label{sec:2D_BG} 2D networks in bilayer graphene}
We begin with the simplest 2D soliton network generated by twists in bilayer graphene. It is well known that pure twists generate screw dislocations. Our goal is to start with a screw dislocation network and see if our inverse design yields pure rotations.

\hfill

\noindent
\underline{\emph{Soliton networks of screw dislocations}}: The defining characteristic of a screw dislocation is that the Burgers vector is parallel to its line direction. However, choosing $\bm l^i$ parallel to $\bm b^i$ in our inverse design framework \emph{will not} result in a pure rotation.\footnote{The resulting heterodeformation will have a non-zero heterostrain, albeit very small.} This is because $\bm l^i$ belongs to the deformed configuration and the dislocation line direction is defined as a \emph{pull-back} of $\bm l^i$. Since the deformation is discontinuous (\cref{eqn:piecewiseF}), the pull-back of $\bm l^i$ is equal to the average of the pull-backs with respect to $\bm F$ and identity: $\frac{1}{2}(\bm F^{-1}+\bm I) \bm l^i$. However, since $\bm F$ is unknown, choosing $\bm l^i$ such that
\begin{equation}
    \bm b^i \, \| \,\frac{1}{2}\left(\bm F^{-1}+\bm I\right) \bm l^i
    \label{eqn:parallel}
\end{equation}
is not possible. In the Section S-2 of Supporting Information, we show that condition $\ref{eqn:parallel}$ is equivalent 
\begin{align}
    &\left (
    \bul l^1- \mathrm{sgn}(\bul l^1 \times \bul l^2) \bul b^2
    \right )\,  \| \, \bul b^1, \quad 
    \left (
    \bul l^2- \mathrm{sgn}(\bul l^2 \times \bul l^3)\bul b^3
    \right ) \, \| \, \bul b^2, \quad 
    \left (
    \bul l^3- \mathrm{sgn}(\bul l^3 \times \bul l^1) \bul b^1
    \right ) \, \| \,\bul b^3.
\label{eqn:lbb}
\end{align}
A general solution satisfying \cref{eqn:lbb} is of the form
\begin{subequations}
\label{eqn:lbbn}
\begin{align}
    &\bul l^1=(k,2k+1), \quad \bul l^2=(k+1,-k) \text{ or } \label{eqn:lbbn1}\\
    &\bul l^1=-(k,2k+1), \quad \bul l^2=(-k-1,k), \quad k\in \mathbb Q \label{eqn:lbbn2}
\end{align}
\end{subequations}
where the latter corresponds to clockwise rotations and the former to counterclockwise rotations.

\begin{figure}[t]
    \centering
        \includegraphics[width=0.5\linewidth]{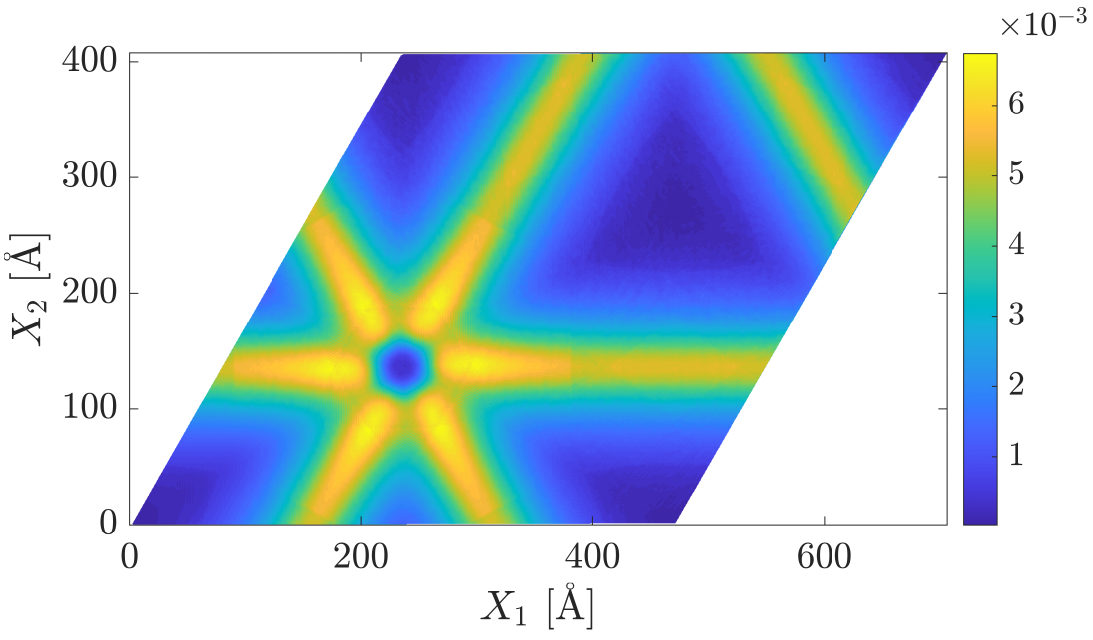}
    \caption{A simple network of screw dislocations in a relaxed $0.299363^\circ$ twisted bilayer graphene.}
    \label{fig:simple_twist}
\end{figure}

Therefore, we proceed with the line directions: $\bul l^1= (-110,-221)$, $\bul l^2= (-111,110)$, and $\bul l^3= -\bul l^1-\bul l^2$, which satisfy the conditions in \cref{eqn:lbbn2}, and Burgers vectors in \cref{eqn:b_graphene}. The resulting transition matrix and the heterodeformation gradient are
\begin{equation}
    \bm Q=\frac{1}{36631}\begin{bmatrix}
        36520 & 221 \\
        -221 & 36741
    \end{bmatrix}, \quad 
    \Fe =\begin{bmatrix}
        0.999986 & -0.005225 \\
        0.005225 & 0.999986
    \end{bmatrix} = \bm R(0.299363^\circ),
\end{equation}
which confirms $\bm F$ is a pure twist. Since all $\bm l^i$ have integer lattice coordinates, we expect the resulting network to be simple. \Cref{fig:simple_twist} shows a relaxed $0.299363^\circ$ twisted bilayer graphene, confirming the presence of screw dislocations in a simple network. Towards the end of this section where we present complex moir\'e, we construct a twist-induced complex moir\'e by choosing a non-integer rational $k$.

\hfill

\noindent
\underline{\emph{Distinct soliton networks with identical translational symmetry}:} Next, we present an exercise, which highlights the one-to-one correspondence between heterodeformations and strain soliton networks. In particular, we construct three distinct heterodeformations of bilayer graphene that have identical moir\'e Bravais lattices but distinct soliton networks.  
\begin{table}[t]
\centering
\small
\renewcommand{\arraystretch}{1.4}
\setlength{\tabcolsep}{8pt}
\begin{tabular}{c c c c}
\hline
 & Network a & Network b & Network c \\
\hline

$\bm \ell^1$ 
& $(1,-280)$ 
& $(1,-280)$ 
& $(1,-280)$ \\
$\bm \ell^2$ 
& $(-281,-1)$ 
& $(-280,-281)$ 
& $(-279,-561)$ \\

$\bm Q$ 
& $\dfrac{1}{78681}
\begin{bmatrix}
78400 & 280 \\
-280 & 78680
\end{bmatrix}$
& $\dfrac{1}{78681}
\begin{bmatrix}
78120 & 279 \\
-280 & 78680
\end{bmatrix}$
& $\dfrac{1}{78681}
\begin{bmatrix}
77840 & 278 \\
-280 & 78680
\end{bmatrix}$ \\

$\bm F$ 
& $\begin{bmatrix}
1.001786 & -0.003093 \\
0.0030929 & 1.001786
\end{bmatrix}$
& $\begin{bmatrix}
1.001792 & -0.003104 \\
0.001020 & 1.005376
\end{bmatrix}$
& $\begin{bmatrix}
1.001799 & -0.003120 \\
-0.001068 & 1.008993
\end{bmatrix}$ \\

$\theta$
&$0.176896^\circ$
&$0.117720^\circ$
&$0.058331^\circ$\\

$\bm U$
&$\begin{bmatrix}
    1.001791 & 0.0 \\
    0.0      & 1.001791
\end{bmatrix}$
&$\begin{bmatrix}
        1.001792 & -0.001038 \\
    -0.001038 & 1.005381
\end{bmatrix}$
&$\begin{bmatrix}
        1.001797 & -0.002088 \\
-0.002088 & 1.008995
\end{bmatrix}$\\

\hline
\end{tabular}
\caption{Three heterodeformations that result in identical moir\'e Bravais lattice but distinct soliton networks.}
\label{tab:moire_degeneracy}
\end{table}

\begin{figure}[t]
    \centering
    \subfloat[]
    {
        \includegraphics[width=0.5\textwidth]{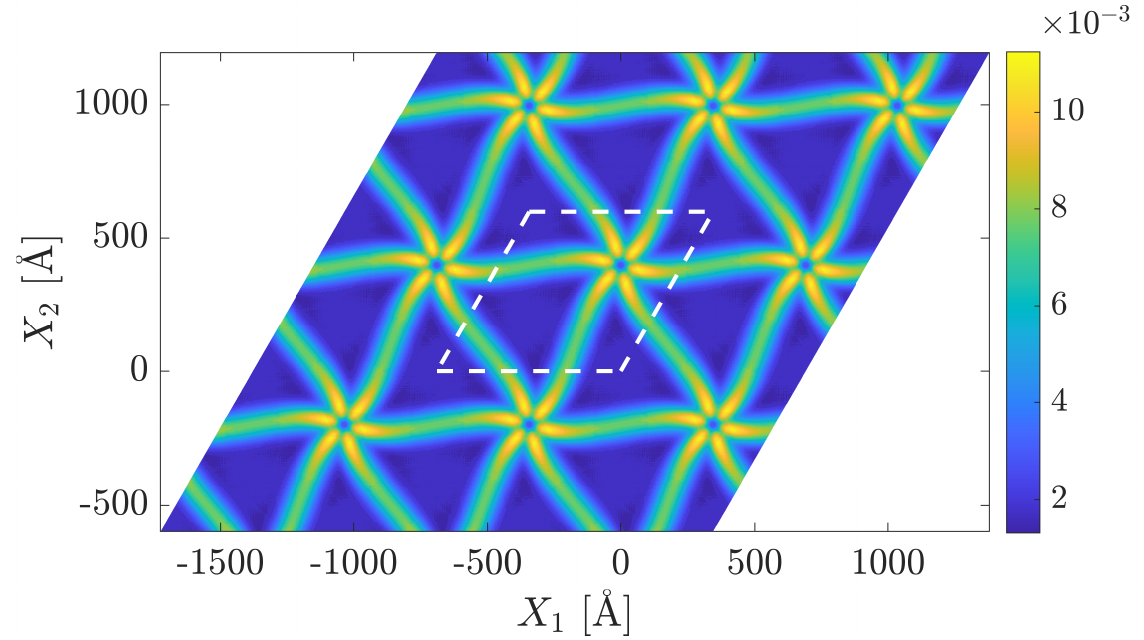}
        \label{fig:triangle_mixed}
    }
    \subfloat[]
    {
    \includegraphics[width=0.5\linewidth]{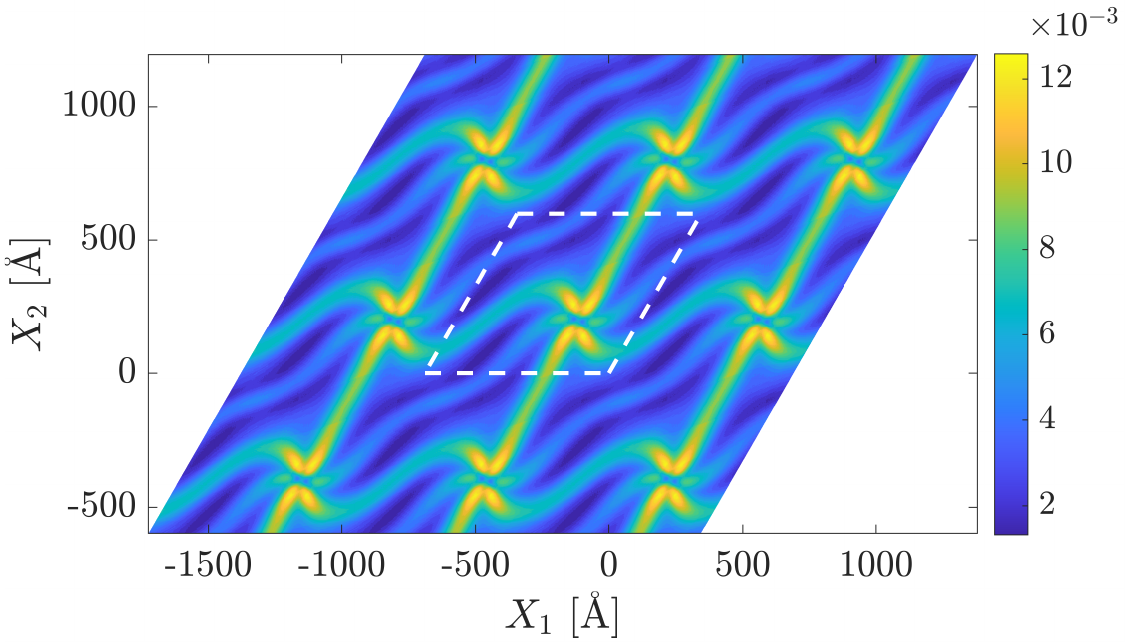}
    \label{fig:triangle_uneq}
    }
    \\
    \subfloat[]
    {
    \includegraphics[width=0.5\linewidth]{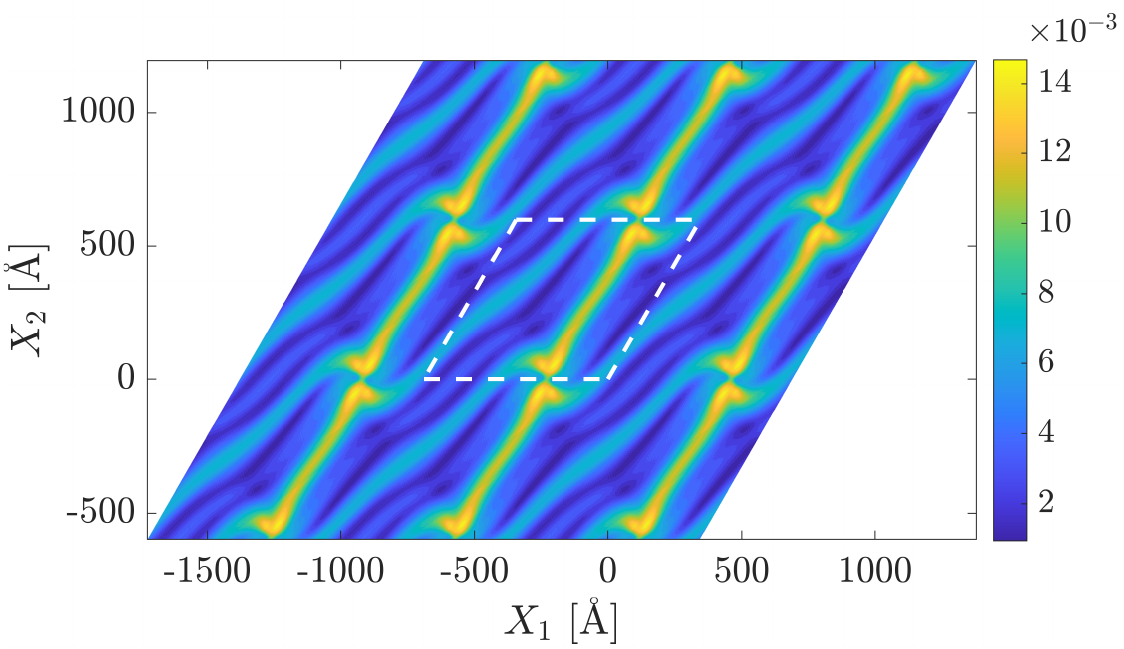}
    \label{fig:mixed_hetero}
    }
    \subfloat[]
    {
    \includegraphics[width=0.4\linewidth]{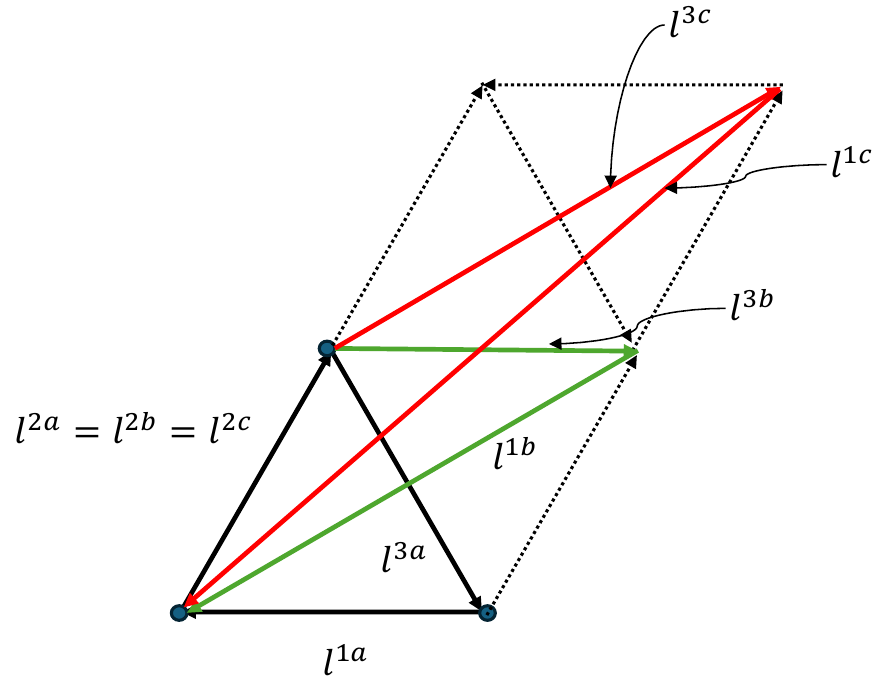}
    \label{fig:degen_summary}
    }
    \caption{
    \psubref{fig:triangle_mixed}-  \psubref{fig:mixed_hetero} Three distinct simple soliton networks in bilayer graphene that have identical moir\'e Bravais lattices (marked in white). The corresponding heterodeformation gradients are listed in \Cref{tab:moire_degeneracy}. \protect\subref{fig:degen_summary} shows the input line vectors of the three soliton networks.} 
    \label{fig:disloc_degenracy}
\end{figure}

For simplicity, we focus on simple networks, by choosing $\bm l^1$ and $\bm l^2$ with integer lattice coordinates. \Cref{tab:moire_degeneracy} includes the line vectors of the three networks, which share the line vector $\bm l^1$. Clearly, the two line vectors of the three networks span the same moir\'e lattice, which is also depicted in \Cref{fig:degen_summary}. \Cref{tab:moire_degeneracy} also includes the corresponding transition matrices, deformation gradients, and their polar decompositions.

\Cref{fig:triangle_mixed,fig:triangle_uneq,fig:mixed_hetero} show the relaxed configurations of the three heterodeformed bilayers. We now make the following observations.
\begin{enumerate}
    \item By design, the interface dislocations in the three bilayers are along their line vectors.
    \item The distributions of the AA stackings are identical across the three bilayers, confirming the shared moir\'e Bravais lattice.
    \item In our earlier work \citep{ahmed2025quantifying,ahmed2025multiscale}, we showed that non-screw dislocations minimize their interaction energy by swirling to increase their screw character. Earlier, spiral solitons were reported \citep{mesple2023giant} in bilayer graphene under equi-biaxial heterostrain. The extent of swirling depends on the length of a dislocation between two adjacent AA pinning points. The dislocations in all three cases are of mixed character and exhibit varying degrees of swirling, as their lengths differ. In \Cref{fig:triangle_mixed}, all three dislocations swirl to the same extent as the orientations of their Burgers vectors relative to the line vectors are identical.
    \item \Cref{fig:triangle_mixed} has the largest point group symmetry ($C_6$), which is broken in \Cref{fig:triangle_uneq,fig:mixed_hetero}. We expect the corresponding continuum Hamiltonians for the flat-band physics of the latter two to reflect this broken symmetry.
\end{enumerate}

\hfill

\noindent
\underline{\emph{Complex networks in bilayer graphene}}: Next, we demonstrate complex 2D networks in bilayer graphene. Using our general solution in \cref{eqn:lbbn} for line vectors that generate pure twists, we construct a pure twist that results in a complex soliton network by choosing a rational $k=-\frac{151}{3}$, leading to $\bul l^1= (-\frac{151}{3},-\frac{305}{3})$, $\bul l^2= (-\frac{154}{3},\frac{151}{3})$, and $\bul l^3= -\bul l^1-\bul l^2$. The resulting transition matrix and deformation gradient are 
\begin{equation}
    \bm Q=\frac{1}{23257}\begin{bmatrix}
        23103 & 305 \\
        -305 & 23408
    \end{bmatrix}, \quad
    \Fe=\begin{bmatrix}
        0.999936 & -0.011357 \\
        0.011357 & 0.999936
    \end{bmatrix}=\bm R(0.650741^\circ),
\end{equation}
which confirms that the heterodeformation is a pure twist. Recall that in the case of complex moir\'e, to construct the bilayer with PBCs, we cannot choose $\bm l^1$ and $\bm l^2$ as the box vectors as the primitive cell of CSL is larger. Therefore, we use SNF bicrystallography to compute the following CSL primitive vectors:
\begin{equation}
    \bm \ell^1 = -375.156047 \,  \bm e_1, \quad \bm \ell^2 =  187.578025 \,  \bm e_1 -324.894669\, \bm e_2.
\end{equation}
\begin{figure}[t]
    \centering
    \subfloat[]{
        \includegraphics[height=0.3\linewidth]{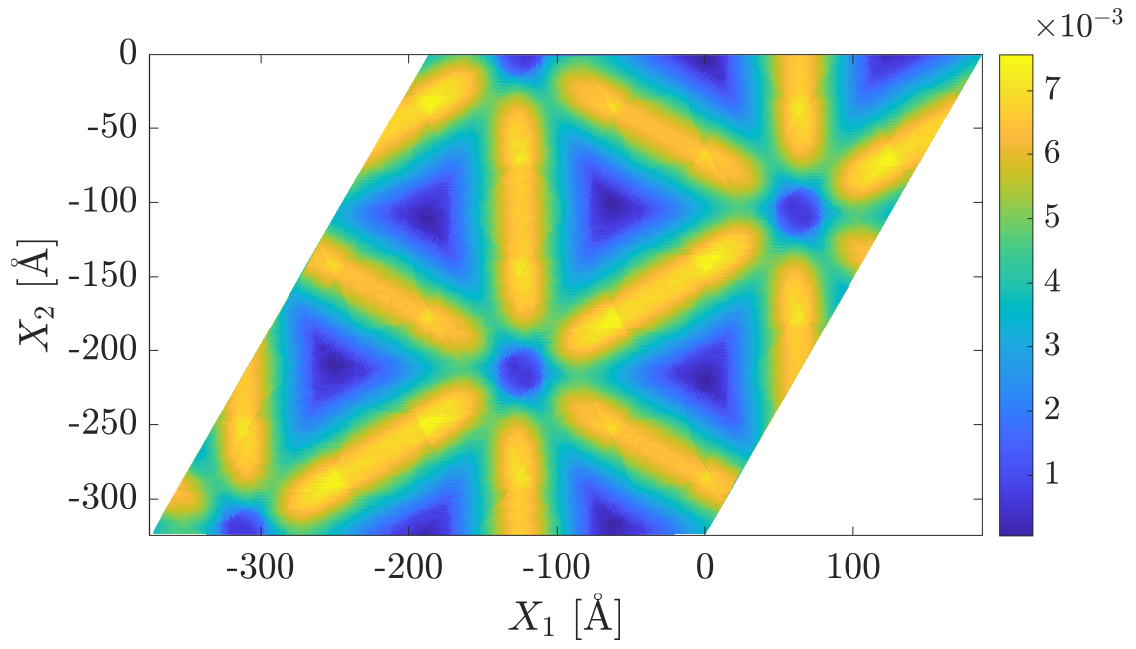}
        \label{fig:complex_twist}
    }
    \subfloat[]{
        \includegraphics[height=0.3\linewidth]{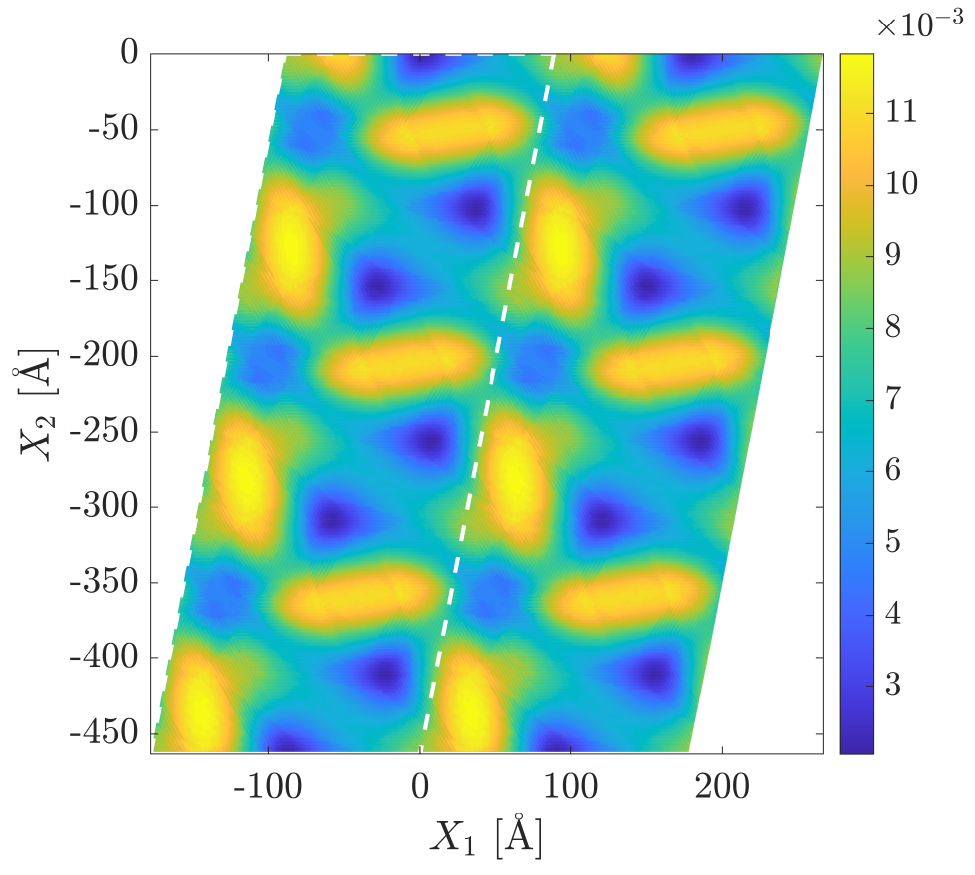}
        \label{fig:complex_moire_2}
    }
    \caption{Complex soliton network in \psubref{fig:complex_twist} $0.650741^\circ$ twisted and \psubref{fig:complex_moire_2} heterodeformed bilayer graphene.}
    \label{fig:complex}
\end{figure}
\Cref{fig:complex_twist} shows a relaxed $0.650741^\circ$ twisted bilayer graphene, simulated using the smallest simulation cell spanned by $\bm \ell^1$ and $\bm \ell^2$. The simulation cell contains $3$ moirons \cite{Hermann_2012} (or moir\'e unit cells).

As a final example of bilayer graphene, we construct a complex moir\'e with one of the line vectors having integer lattice coordinates while the other two have rational lattice coordinates:
$\bul l^1= (1,-72)$, $\bul l^2= (-218/3,-25)$ and, $\bul l^3= (215/3,97)$. The resulting transition matrix and deformation gradient are 
\begin{align*}
    \bm Q&=\frac{1}{15771}\begin{bmatrix}
       15480 & 215 \\
        -216 & 15768
    \end{bmatrix}, \\
    \Fe&=\begin{bmatrix}
        1.006977 & -0.012084 \\
        0.009324 & 1.011628
        \label{eq:complex_moire_2}
    \end{bmatrix}=\bm R(0.607625^\circ) 
    \begin{bmatrix}
        1.007019 & -0.001355 \\
    -0.001355 & 1.011699
    \end{bmatrix}.
\end{align*}
SNF bicrystallography yields the following simulation box vectors:
\begin{equation}
    \bm \ell^1 = -178.362723 \,  \bm e_1, \quad \bm \ell^2 =  89.181354 \,  \bm e_1 -463.399954\, \bm e_2.
\end{equation}

\Cref{fig:complex_moire_2} shows the relaxed heterodeformed configuration. For better visualization, we show two replicas of the simulation cell along the $X_1$ direction. The simulation domain contains three moirons/moir\'e unit cells.

\subsubsection{\label{sec:2D_simple_mos2} 2D networks in bilayer MoS$_2$}
In this section, we demonstrate inverse design in MoS$_2$ and highlight key differences from bilayer graphene. Inspired by the second exercise in \Cref{sec:2D_BG}, we construct two distinct honeycomb soliton networks that share the same moir\'e Bravais lattice.

\begin{figure}[t]
    \centering
    \subfloat[]
    {
        \includegraphics[width=0.5\textwidth]{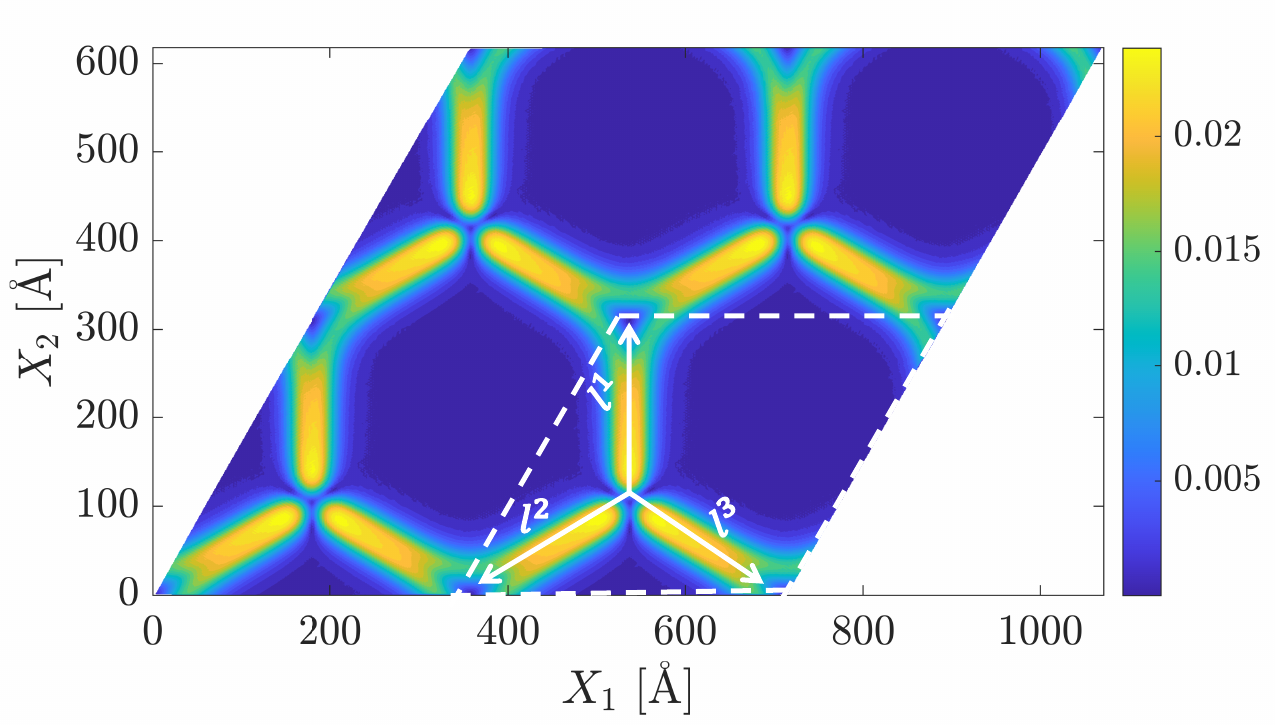}
        \label{fig:mos2_degen_a1}
    }
    \subfloat[]
    {
    \includegraphics[width=0.5\linewidth]{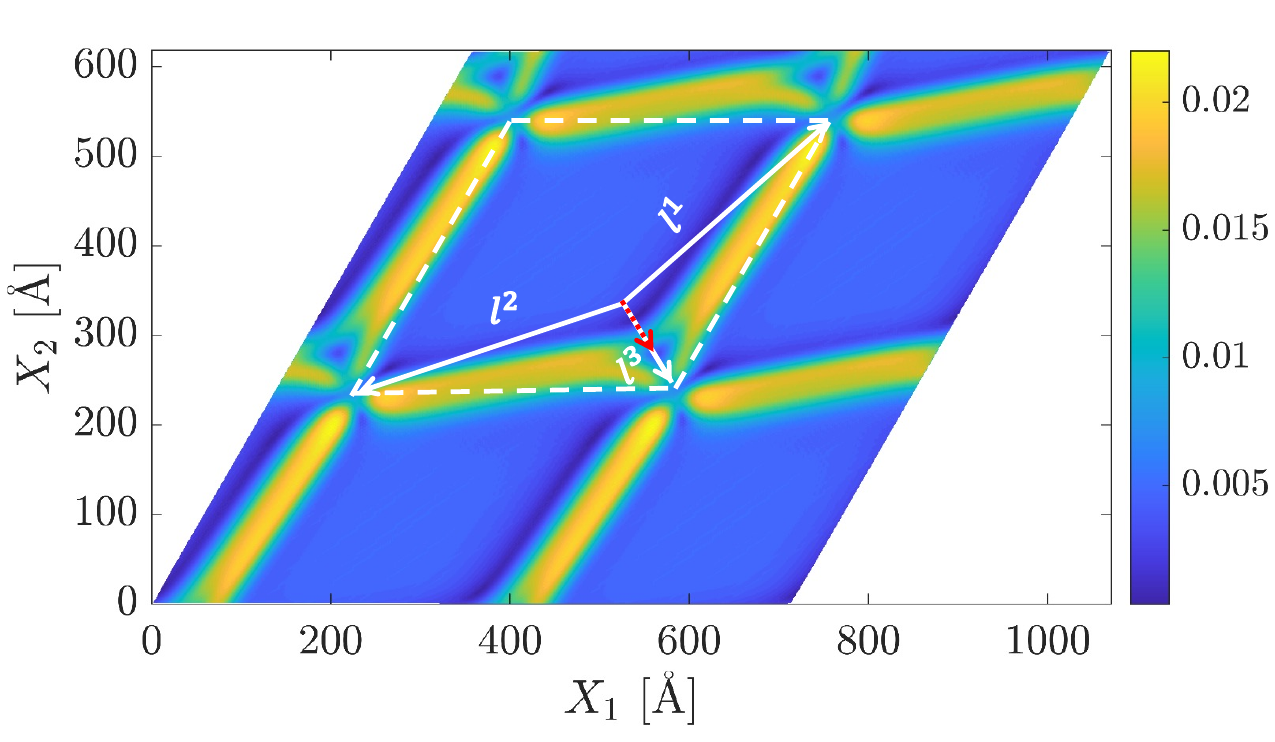}
    \label{fig:mos2_degen_a2_relax}
    }
    \caption{
    \psubref{fig:mos2_degen_a1},  \psubref{fig:mos2_degen_a2_relax} Two distinct simple soliton networks in bilayer MoS$_2$ that have identical moir\'e Bravais lattices (marked in white). The corresponding heterodeformation gradients are listed in \cref{eq:eq_hetero_mos2,eq:distort_hetero_mos2}. The soliton lines in the relaxed structure in \psubref{fig:mos2_degen_a2_relax} do not exactly align with the input line vectors (marked as white arrows) as the heterodeformation does not uniquely determine the line directions.}
    \label{fig:mos2_disloc_degenracy}
\end{figure}
For the first network, we pick line directions
$\bul l^{1a}= (-64,1)$ and $\bul l^{2a}= (65,64)$, resulting in 
\begin{align}
    \bm Q&=\frac{1}{4161}\begin{bmatrix}
        4140 & 43 \\
        -43 & 4183
    \end{bmatrix}, \notag\\
    \Fe&=\begin{bmatrix}
        0.999800 & -0.008947 \\
        0.008947 & 0.999800 
    \end{bmatrix}=
    \bm R(0.512696^\circ) 
    \begin{bmatrix}
        0.999840 & 0 \\
        0 & 0.999840
    \end{bmatrix}.
    \label{eq:eq_hetero_mos2}
\end{align}
\Cref{fig:mos2_degen_a1} shows the relaxed configuration corresponding to \cref{eq:eq_hetero_mos2} containing a regular hexagonal soliton network along the chosen line directions. The dashed parallelogram highlights the moir\'e unit cell.

For the second network, the line directions are chosen such that they form the same moir\'e unit cell as in the previous network:  $\bul l^{1b}= (-108,-85)$, $\bul l^{2b}= (87,107)$ and, $\bul l^{3b}= -\bul l^{1b}-\bul l^{2b}$. The resulting transition matrix and deformation gradient are
\begin{align}
    \bm Q&=\frac{1}{4161}\begin{bmatrix}
        4097 & 65 \\
        -43 & 4183
    \end{bmatrix},\notag \\
    \Fe&=
    \begin{bmatrix}
        0.999798 & -0.009040 \\
        0.009180 & 1.010236 
    \end{bmatrix}=\bm R(0.519354^\circ) 
    \begin{bmatrix}
        0.999840 & 0.000117 \\
        0.000117 & 1.010277
    \end{bmatrix}.
    \label{eq:distort_hetero_mos2}
\end{align}
The relaxed configuration corresponding to \cref{eq:distort_hetero_mos2}, shown in \Cref{fig:mos2_degen_a2_relax}, contains a network of distorted hexagons. By design, the moir\'e unit cell (dashed parallelogram) is identical to that in \Cref{fig:mos2_degen_a1}. However, the three dislocation lines are not exactly in alignment with the input dislocation line vectors (white arrows). This is because, in the case of MoS$_2$, the heterodeformation does not uniquely determine the line directions,\footnote{Recall that the condition in \cref{eqn:lsum} is always satisfied in bilayer graphene but may not hold in MoS$_2$, and it was imposed to make the inverse design problem well-posed.} whose orientation in the relaxed configuration ultimately depends on the energetics of the dislocations.

As a last example, we demonstrate a complex network in bilayer MoS$_2$ using line vectors $\bul l^{1}= (-\frac{216}{5},\frac{1}{5})$ and $\bul l^{2}= (\frac{217}{5},\frac{216}{5})$. For this set of line vectors, and Burgers vectors in \cref{eqn:b_mos2}, the transition matrix is 
\begin{equation}
    \bm Q=\frac{1}{140619}\begin{bmatrix}
        139544 & 2165 \\
        -2165 & 141709
    \end{bmatrix},
\end{equation}
and the deformation gradient is
\begin{equation}
    \Fe=\begin{bmatrix}
    0.999769 & -0.013330 \\
        0.013330 & 0.999769 
    \end{bmatrix}=\bm R(0.763868^\circ) 
    \begin{bmatrix}
        0.999858 & 0 \\
    0 & 0.999858
    \end{bmatrix}.
    \label{eqn:complex_mos2_F}
\end{equation}
SNF bicrystallography yields the following box vectors for the smallest simulation cell:
\begin{equation}
    \bm \ell^1 = 1198.856929 \,  \bm e_1, \quad \bm \ell^2 =  599.428459 \,  \bm e_1 +1038.2405476\, \bm e_2.
\end{equation}
The relaxed configuration corresponding to the deformation in \cref{eqn:complex_mos2_F} is shown in \Cref{fig:mos2_degen_a1}. The simulation cell contains $25$ primitive moir\'e unit cells. In addition, the figure highlights the inequivalence of A'B and AB' stackings, which is exacerbated under this heterodeformation. 
\begin{figure}
    \centering
    \includegraphics[width=0.5\linewidth]{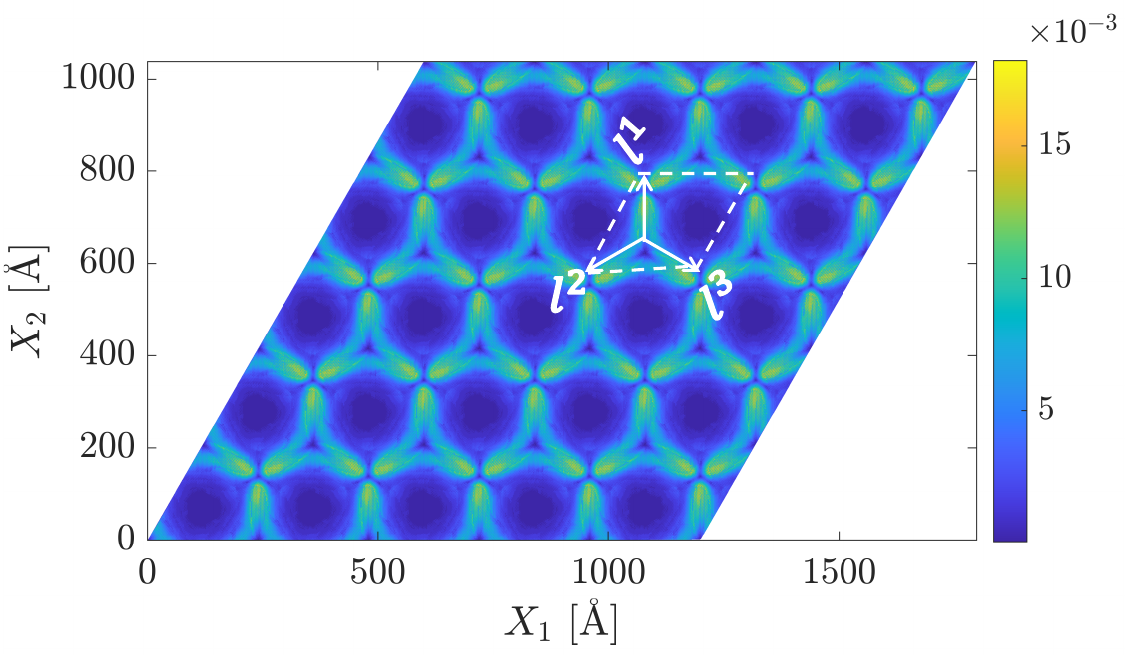}
    \caption{A complex soliton network in bilayer MoS$_2$.}
    \label{fig:mos2_complex}
\end{figure}

The examples presented in this section demonstrate the generality of the developed inverse design framework for constructing heterodeformations that yield the desired soliton network in a wide range of  van der Waals homostructures.

\section{Summary and conclusions}
\label{sec:conclusion_inverse}
We developed a geometric framework that establishes a one-to-one correspondence between heterodeformations and strain soliton networks in bilayer 2D materials. The central idea is to represent the network through line direction--Burgers vector pairs, which enables the heterodeformation to be computed directly from the network geometry. The admissible line directions are constrained by the topology of the soliton network (e.g., triangular networks in bilayer graphene and hexagonal networks in MoS$_2$), which is dictated by the symmetries of the GSFE extrema. In this way, the GSFE enters the construction through topological constraints, ensuring that the resulting heterodeformation is consistent with the underlying energetics. This enables an inverse design strategy: given a target soliton network, one can construct the heterodeformation and the periodic simulation cell that yield the desired network upon structural relaxation.

Most existing approaches to moir\'e systems adopt a forward construction in which the heterodeformation is specified and the resulting moir\'e Bravais lattice is analyzed. In contrast, the present framework adopts an inverse perspective in which the target soliton network is prescribed and the corresponding heterodeformation is constructed. From this viewpoint, the moir\'e Bravais lattice alone is insufficient to characterize the interface: distinct heterodeformations can share the same moir\'e lattice while producing different soliton networks. This reflects an inherent many-to-one mapping when only translational symmetry is retained, and highlights the need to work at the level of the full multilattice structure in inverse design.

The proposed framework is general and applies to arbitrary bilayer systems. It enables the systematic construction of interfacial structures with prescribed topology and symmetry, providing a practical route for exploring the space of moir\'e interfaces beyond conventional twist-based configurations and for engineering networks tailored to specific physical properties.

More broadly, the framework establishes a geometric foundation for connecting GSFE landscapes, network topology, and heterodeformation, and can be integrated with continuum and atomistic models to study structure--property relations in moir\'e materials. This opens the door to the rational design of moir\'e systems with controlled structural and electronic characteristics.

\backmatter

\bmhead{Acknowledgements}
NCA and TA would like to acknowledge support from the National Science Foundation Grant NSF-MOMS-2239734 with S. Qidwai as the program manager.

\section*{Supplementary information}
Supplementary information accompanies this paper.

\section*{Declarations}
\begin{itemize}
\item Data availability: Initial and final configurations from the MD simulations are 
publicly available on Materials Cloud at \url{https://doi.org/10.24435/materialscloud:ar-da}. Additional data supporting the findings of this study are available from the corresponding author upon reasonable request.
\item Code availability: The implementation of the proposed framework is available as an open-source C++ library (open Interface Lab, \texttt{oILAB}) at \url{https://github.com/admal-research-group/oILAB.git}, with Python bindings for user-level workflows. Reproducible examples demonstrating the inverse design of soliton networks are provided as a Python script in the \texttt{heterodeformation/} directory.
\item Competing interest: The authors declare no competing financial or non-financial interests.
\item Author contribution: N.A. conceived the research, developed the theoretical framework, and performed the derivations. T.A. implemented the computational framework and carried out all calculations. N.A. and T.A. analyzed the results and wrote the manuscript.
\end{itemize}

\bibliography{achemso-demo}

\end{document}